\documentclass[a4paper]{article}
\usepackage[margin=25mm]{geometry}
\usepackage{amsmath}
\usepackage{amsfonts}
\usepackage{amssymb}
\usepackage{amsthm}
\usepackage[dvips]{graphicx}
\usepackage{authblk}
\usepackage{cite}

\providecommand{\keywords}[1]
{
  \small	
  \textbf{Keywords:} #1
}

\newcommand{\be}{\begin{equation}}
\newcommand{\ee}{\end{equation}}

\numberwithin{equation}{section}

\begin{document}

\title{The Husimi function of a semiconfined harmonic oscillator model with a position-dependent effective mass}

\author{E.I. Jafarov\thanks{Corresponding author: ejafarov@physics.science.az}}
\author{A.M. Jafarova\thanks{a.jafarova@physics.science.az}}
\author{S.M. Nagiyev\thanks{smnagiyev@physics.ab.az}}

\affil{Institute of Physics, Azerbaijan National Academy of Sciences, Javid av. 131, AZ1143, Baku, Azerbaijan}

\date{} 

\maketitle

\begin{abstract}

The phase space representation for a semiconfined harmonic oscillator model with a position-dependent effective mass is constructed. We have found the Husimi distribution function for the stationary states of the oscillator model under consideration for both cases without and with the applied external homogeneous field. The obtained function is expressed through the double sum of the parabolic cylinder function. Different special cases and the limit relations are discussed, too.

\end{abstract}


\section{Introduction}

Exactly solvable problems in quantum mechanics are always considered the most attractive for scientists belonging to different branches. The reason for their attractivity is that the exact solution within the fundamental principles of the quantum approach makes their use easier for further successful applications, both theoretical and experimental. The problem of the quantum harmonic oscillator having an exact solution within the non-relativistic canonical approach is one of them. Its wavefunctions of the stationary states are the eigenfunctions of the second-order differential Schr\"odinger equation. The polynomial part of the analytical expression of these eigenfunctions is expressed via the Hermite polynomials. The square root of their weight function is of the Gaussian behavior and is strictly responsible for the vanishing of the wave function at $x\to \pm \infty$. The energy spectrum of the quantum harmonic oscillator problem, obtained from the mentioned above second-order differential equation, consists of an infinite number of equidistant energy levels.

Recently, a new model of the exactly-solvable one-dimensional non-relativistic canonical quantum harmonic oscillator has been presented~\cite{jafarov2021}. The model exhibits a semiconfinement effect, i.e. its wavefunctions of the stationary states vanish at both values of the position $x=-a$ and $x \to +\infty$. In that model, the semiconfinement effect is achieved by replacing the constant effective mass with a position-dependent effective mass. Although its wavefunctions are expressed using generalized Laguerre polynomials, surprisingly, its energy spectrum completely coincides with that of the non-relativistic canonical quantum harmonic oscillator described above. Later, the same quantum oscillator model was generalized to the case of the applied external homogeneous field~\cite{jafarov2022}. We observed that the wavefunctions of the oscillator model generalized via the applied external homogeneous field are still expressed through the generalized Laguerre polynomials, its energy spectrum is still equidistant but depends on the semiconfinement parameter $a$. When we depicted the probability densities of its wavefunctions in the configuration representation, then one could observe that the semiconfined quantum system under construction stays close to the infinite high wall with higher probability. Even, the applied external homogeneous field does not change such behavior of it significantly. Taking this phenomenon into account and in order to understand better the general behavior of the oscillator model exhibiting semiconfinement effect, we decided to apply the most powerful tool -- to construct its phase space and to explore its behavior via the exact expression of the Gaussian smoothed Wigner function of the joint quasiprobability of the momentum and position.

We structured the rest of the paper as follows: Section 2 consists of the known information about the definition of the Gaussian smoothed Wigner function and provides the analytical expression of it for the standard nonrelativistic canonical quantum harmonic oscillator -- both cases without and with the applied external homogeneous field. The next section is devoted to the computation of the Gaussian smoothed Wigner function for the oscillator model with a position-dependent effective mass exhibiting semiconfinement effect -- computations are done also for both cases without and with the applied external homogeneous field. The final section is devoted to the detailed discussions and includes also a brief conclusion.

\section{The Wigner distribution function and its Gaussian smoothing}

The Wigner distribution function is the powerful tool of quantum physics that allows us to describe the quantum systems under study by using the language of classical physics. It was first introduced empirically in~\cite{wigner1932} as a tool to study possible quantum corrections to the thermodynamic equilibrium. If one considers the quantum system of the stationary behavior, then, the Wigner function of the stationary states corresponding to such a system can be obtained by using its wavefunctions of the stationary states as follows:

\be
\label{wf-gen}
W_n \left( {p,x} \right) = \frac{1}{{2\pi \hbar }}\int { \psi _n^* \left( {x - \frac{1}{2}x'} \right) \psi _n \left( {x + \frac{1}{2}x'} \right)e^{ - i\frac{{px'}}{\hbar }} dx'} ,
\ee
where $ \psi _n \left( x \right)$ is an analytical expression of the orthonormalized wavefunction of the stationary states of the quantum system under consideration in the position representation. The same analytical expression for the Wigner distribution function can be obtained via its similar definition by employing the wavefunctions of the stationary states in the momentum representation.  The distribution function defined as (\ref{wf-gen}) is strictly positive only for the wavefunctions of the Gaussian behavior. Generally, it is bounded via the restriction $\left| {W_n \left( {p,x} \right)} \right| \le \left( {\pi \hbar } \right)^{ - 1} $, i.e., it can take both positive and negative values. Due to such a property, it is called a joint quasiprobability distribution function of momentum $p$ and position $x$. Then, in order to achieve a positivity behavior for the joint quasiprobability distribution function of momentum $p$ and position $x$ defined as (\ref{wf-gen}), Gaussian smoothing for it by the following manner can be applied~\cite{hillery1984}:

\be
\label{gs-wf}
\bar W_n \left( {p,x} \right) = \frac{1}{{\pi \hbar }}\int\limits_{ - \infty }^\infty  {\int\limits_{ - \infty }^\infty  {e^{ - \frac{{p'^2 }}{{2\Delta _p ^2 }} - \frac{{x'^2 }}{{2\Delta _x ^2 }}} W_n \left( {p + p',x + x'} \right)dp'dx'} }.
\ee

Gaussian smoothed Wigner function (\ref{gs-wf}) is always positive for the ($\Delta _p \Delta _x$) finite region of the phase plane, i.e. $\bar W_n \left( {p,x} \right)\geq 0$. Simplest realization of the Gaussian smoothed Wigner function in the finite region of the phase plane corresponds to the case $\Delta _p \Delta _x=\hbar/2$.  It leads to the definition of the Husimi distribution function with a normalization factor $\left( {\pi \hbar } \right)^{ - 1} $ that allows to compare its behavior with definition (\ref{wf-gen})~\cite{husimi1940}:

\be
\label{hf-gen}
\bar W_n \left( {p,x} \right) = \frac{1}{{\left( {2\pi } \right)^{\frac{3}{2}} \hbar \Delta _x }}\left| {\int { \psi _n \left( {x'} \right)e^{ - i\frac{{px'}}{\hbar } - \frac{{\left( {x - x'} \right)^2 }}{{4\Delta _x ^2 }}} dx'} } \right|^2 .
\ee

It is restricted as $0 \le \bar W_n \left( {p,x} \right)\le \left( {\pi \hbar } \right)^{ - 1}$. Further, the value of the parameter $\Delta _x^2$ taken equal to $\hbar/2 m_0 \omega$ allows to compute the exact expression of the Husimi function (\ref{hf-gen}) of the quantum system under consideration by employing the analytical expression of its wavefunctions of the stationary states. For example, the Husimi function of the non-relativistic quantum harmonic oscillator under the action of the external homogeneous field $V^{ext}\left(x \right)=gx$ can be exactly computed via its following orthonormalized wavefunctions of the stationary states

\be
\label{wf-gh}
\psi _{Nn}^g \left( x \right) = C_{Nn} e^{ - \frac{{\lambda _0 ^2 }}{2}\left( {x + x_0 } \right)^2 } H_n \left( {\lambda _0 \left( {x + x_0 } \right)} \right). 
\ee

Here, $H_n \left( x \right)$ is the Hermite polynomial defined via the $_2F_0$ hypergeometric functions~\cite{koekoek2010} and the following notations are introduced, too:

\[
\lambda _0  = \sqrt {\frac{{m_0 \omega }}{\hbar }} ,\quad x_0  = \frac{g}{{m_0 \omega ^2 }},\quad C_{Nn}  = \frac{{C_{N0} }}{{\sqrt {2^n n!} }},\quad C_{N0}  = \sqrt[4]{{\frac{{\lambda _0 ^2 }}{\pi }}}.
\]

We are not aware that the analytical expression of the Husimi function computed via (\ref{wf-gh}) exists in scientific literature. However, it does not drastically differ from the Husimi function computed for the non-relativistic quantum harmonic oscillator without the action of the external homogeneous field. Below, we show how the function itself can be computed via (\ref{wf-gh}) by employing two methods.

Taking into account that the wavefunction possesses an orthogonality relation within the region $\left(-\infty,+\infty \right)$, then the Husimi function of the non-relativistic quantum harmonic oscillator under the action of the external homogeneous field being computed via (\ref{wf-gh}) will have the following general form:

\be
\label{hf-ogh}
\bar W_n \left( {p,x} \right) \equiv \bar W_{Nn}^g \left( {p,x} \right) = \frac{{\lambda _0 }}{{2\pi \hbar \sqrt \pi  }}\left| {Q_{Nn}^g } \right|^2 ,
\ee
where,

\be
\label{qnng-1}
Q_{Nn}^g  = \int\limits_{ - \infty }^\infty  {\psi _{Nn}^g \left( {x'} \right)e^{ - i\frac{{px'}}{\hbar } - \frac{{\lambda _0 ^2 }}{2}\left( {x - x'} \right)^2 } dx'} .
\ee

Substitution of (\ref{wf-gh}) at (\ref{qnng-1}) yields

\be
\label{qnng-2}
Q_{Nn}^g  = C_{Nn} \int\limits_{ - \infty }^\infty  {H_n \left( {\lambda _0 \left( {x' + x_0 } \right)} \right) \cdot e^{ - i\frac{{px'}}{\hbar } - \frac{{\lambda _0 ^2 }}{2}\left( {x - x'} \right)^2 } e^{ - \frac{{\lambda _0 ^2 }}{2}\left( {x' + x_0 } \right)^2 } dx'} .
\ee

The first method of its computation is the so-called operator method. Let's take into account that

\[
i\hbar \frac{d}{{dp}}e^{ - i\frac{{px'}}{\hbar }}  = i\hbar \left( { - i\frac{{x'}}{\hbar }} \right)e^{ - i\frac{{px'}}{\hbar }}  = x'e^{ - i\frac{{px'}}{\hbar }} .
\]

Then, one can write down

\be
\label{hp-1}
H_n \left( {\lambda _0 \left( {x' + x_0 } \right)} \right) \cdot e^{ - i\frac{{px'}}{\hbar }} = H_n \left( {i\hbar \lambda _0 \frac{d}{{dp}} + \lambda _0 x_0 } \right) \cdot e^{ - i\frac{{px'}}{\hbar }} .
\ee

As a consequence of (\ref{hp-1}), $Q_{Nn}^g$ defined through (\ref{qnng-2}) also becomes as simple as follows:

\be
\label{qnng-3}
Q_{Nn}^g  = \frac{1}{{\sqrt {2^n n!} }}H_n \left( {i\hbar \lambda _0 \frac{d}{{dp}} + \lambda _0 x_0 } \right) Q_{N0}^g ,
\ee
where,

\[
Q_{N0}^g  = C_{N0} \int\limits_{ - \infty }^\infty  {e^{ - \frac{{\lambda _0 ^2 }}{2}\left( {x' + x_0 } \right)^2  - \frac{{\lambda _0 ^2 }}{2}\left( {x - x'} \right)^2  - i\frac{{px'}}{\hbar }} dx'} .
\]

$Q_{N0}^g$ can be easily computed by using the following well-known integral of a Gaussian function:

\[
\int\limits_{ - \infty }^\infty  {e^{ - px^2  - qx} dx}  = \sqrt {\frac{\pi }{p}} e^{\frac{{q^2 }}{{4p}}} ,\quad {\mathop{\rm Re}\nolimits} p > 0.
\]

It equals to

\be
\label{qn0g}
Q_{N0}^g  = \frac{1}{{C_{N0} }}e^{\varphi \left( \eta  \right)} ,
\ee
where,

\be
\label{phi-eta}
\varphi \left( \eta  \right) =  - \frac{{\Delta ^2  + \eta ^2 }}{4} + i\frac{1}{2}\delta \eta , \qquad \Delta  = \xi _0  + \xi ,\quad \delta  = \xi _0  - \xi .
\ee

Here, we also introduced the following dimensionless variables and parameters:

\be
\label{l-0}
\xi  = \lambda _0 x,\quad \xi _0  = \lambda _0 x_0 ,\quad \eta  = \frac{p}{{\sqrt {m_0 \omega \hbar } }} = \frac{p}{{\hbar \lambda _0 }},\quad \frac{d}{{dp}} = \frac{1}{{\hbar \lambda _0 }}\frac{d}{{d\eta }}.
\ee

As a consequence of (\ref{l-0}), the following transfer holds for the differential operator

\be
\label{h-op}
H_n \left( {i\hbar \lambda _0 \frac{d}{{dp}} + \lambda _0 x_0 } \right) \equiv H_n \left( {i\frac{d}{{d\eta }} + \xi _0 } \right),
\ee
and $Q_{N0}^g$ itself:

\be
\label{qnng-4}
Q_{Nn}^g  = \frac{1}{{C_{N0} \sqrt {2^n n!} }}H_n \left( {i\frac{d}{{d\eta }} + \xi _0 } \right)e^{\varphi \left( \eta  \right)} .
\ee

Let's compute the action of operator (\ref{h-op}) to $e^{\varphi \left( \eta  \right)}$. Therefore, we introduce in (\ref{qnng-4}) new function $Z_n \left( \eta  \right)$ as follows:

\be
\label{z_n}
Z_n \left( \eta  \right) = H_n \left( {i\frac{d}{{d\eta }} + \xi _0 } \right)e^{\varphi \left( \eta  \right)} .
\ee

We will use the following generating function for Hermite polynomials~\cite{koekoek2010}:

\[
\sum\limits_{n = 0}^\infty  {\frac{{z^n }}{{n!}}H_n \left( x \right)}  = e^{2xz - z^2 } .
\]

Then, one observes that

\be
\label{i-t0}
I\left( t \right) = \sum\limits_{n = 0}^\infty  {\frac{{t^n }}{{n!}}Z_n \left( \eta  \right)}  = e^{2t\xi _0  - t^2 } e^{2it\frac{d}{{d\eta }}} e^{\varphi \left( \eta  \right)} .
\ee

Taking into account known property of the action of any finite-difference operator $\exp \left( {\alpha \frac{d}{{dx}}} \right)f\left( x \right) = f\left( {x + a} \right)$, one has

\be
\label{i-t}
I\left( t \right) = e^{2t\xi _0  - t^2 } e^{\varphi \left( {\eta  + 2it} \right)} .
\ee

Further trivial computations by employing (\ref{phi-eta}) yield

\[
\varphi \left( {\eta  + 2it} \right) = \varphi \left( \eta  \right)  - t\left( {\delta  + i\eta } \right) + t^2.
\]

Its substitution at (\ref{i-t}) leads to

\be
\label{i-t2}
I\left( t \right) = e^{t\left( {\Delta  - i\eta } \right)} e^{\varphi \left( \eta  \right)} .
\ee

From (\ref{i-t0}) and (\ref{i-t2}) it is clear that

\[
Z_n \left( \eta  \right) = \frac{{d^n }}{{dt^n }}I\left( t \right)|_{t = 0}  = \left( {\Delta  - i\eta } \right)^n e^{\varphi \left( \eta  \right)} .
\]

Substitution of $Z_n \left( \eta  \right)$ at (\ref{qnng-4}) yields

\be
\label{qnng-5}
Q_{Nn}^g  = \frac{1}{{C_{N0} \sqrt {2^n n!} }}\left( {\Delta  - i\eta } \right)^n e^{ - \frac{{\Delta ^2  + \eta ^2 }}{4} + i\frac{1}{2}\delta \eta } .
\ee

Finally, the substitution of analytical expression of $Q_{Nn}^g$ defined by (\ref{qnng-5}) at (\ref{hf-ogh}) allows to compute exact expression of the Husimi function as follows:

\[
\bar W_{Nn}^g \left( {p,x} \right) = \frac{1}{{2\pi \hbar n!}}\left( {\frac{{\left| {\Delta  - i\eta } \right|^2 }}{2}} \right)^n e^{ - \frac{{\Delta ^2  + \eta ^2 }}{2}} .
\]

Now, recovering momentum $p$ and position $x$ from the dimensionless variables $\Delta$ and $\eta$ via (\ref{phi-eta}), one obtains for the Husimi function

\be
\label{hf-gh}
\bar W_{Nn}^g \left( {p,x} \right) = \frac{1}{{2\pi \hbar n!}}\left[ {\frac{1}{{\hbar \omega }}\left( {\frac{{p^2 }}{{2m_0 }} + \frac{{m_0 \omega ^2 }}{2}x^2  + gx + \frac{{g^2 }}{{m_0 \omega ^2 }}} \right)} \right]^n e^{ - \frac{1}{{\hbar \omega }}\left( {\frac{{p^2 }}{{2m_0 }} + \frac{{m_0 \omega ^2 }}{2}x^2  + gx + \frac{{g^2 }}{{m_0 \omega ^2 }}} \right)}. 
\ee

Its analytical expression for the special case of the absence of the external field ($g=0$) is also well known~\cite{tatarskii1983}:

\be
\label{hf-h}
\bar W_{Nn}^0 \left( {p,x} \right) \equiv \bar W_n^0 \left( {p,x} \right) = \frac{1}{{2\pi \hbar n!}}\left[ {\frac{1}{{\hbar \omega }}\left( {\frac{{p^2 }}{{2m_0 }} + \frac{{m_0 \omega ^2 }}{2}x^2 } \right)} \right]^n e^{ - \frac{1}{{\hbar \omega }}\left( {\frac{{p^2 }}{{2m_0 }} + \frac{{m_0 \omega ^2 }}{2}x^2 } \right)} .
\ee

This analytical expression also can be computed directly by employing the analytical expression of its wavefunctions of the stationary states of the non-relativistic quantum harmonic oscillator~\cite{landau1991}:

\be
\label{wf-h}
\psi _{Nn}^0 \left( x \right) = \frac{1}{{\sqrt {2^n n!} }}\left( {\frac{{\lambda _0 ^2 }}{\pi }} \right)^{\frac{1}{4}} e^{ - \frac{{\lambda _0 ^2 }}{2}x^2 } H_n \left( {\lambda _0 x} \right).
\ee

We already noted that the analytical expression (\ref{hf-gh}) of the Husimi function can be computed by employing two different methods. The first method was the so-called operator method. Actually, two different methods are applicable for the computation of integral (\ref{qnng-2}). Then, the substitution of its square of absolute value at (\ref{hf-ogh}) gives the analytical expression of the Husimi function. Let's perform the computation of integral (\ref{qnng-2}) by employing the second method -- it is the so-called direct computation method. Simplification of the exponential function from the integrand of (\ref{qnng-2}) allows to rewrite $Q_{Nn}^g$ as follows:

\be
\label{qnng-6}
Q_{Nn}^g  = C_{Nn} \int\limits_{ - \infty }^\infty  {e^{ - \lambda _0 ^2 x'^2  + \beta _N x' + \beta _{N0} } H_n \left( {\lambda _0 \left( {x' + x_0 } \right)} \right)dx'} .
\ee

Here,

\[
\beta _N  = \lambda _0 ^2 \left( {x - x_0 } \right) - i\frac{p}{\hbar },\quad \beta _{N0}  =  - \frac{{\lambda _0 ^2 }}{2}\left( {x^2  + x_0 ^2 } \right).
\]

Now, introducing a new dimensionless variable $y = \lambda _0 \left( {x' + x_0 } \right)$, we obtain that

\be
\label{qnng-7}
Q_{Nn}^g  = \frac{{C_{Nn} }}{{\lambda _0 }}\int\limits_{ - \infty }^\infty  {e^{ - y^2  + \left( {\Delta  - i\eta } \right)y + \beta _0 } H_n \left( y \right)dy},
\ee
where $\beta _0  =  - \frac{1}{2}\Delta ^2  + i\xi _0 \eta$.

Further trivial mathematical tricks on the exponential function of the integrand allow to rewrite eq.(\ref{qnng-7}) as follows:

\be
\label{qnng-8}
Q_{Nn}^g  = \frac{{C_{Nn} }}{{\lambda _0 }}e^{\frac{1}{4}\beta ^2  + \beta _0 } \int\limits_{ - \infty }^\infty  {e^{ - \left( {\frac{1}{2}\beta  - y} \right)^2 } H_n \left( y \right)dy} ,
\ee
where $\beta  = \Delta  - i\eta$.

One observes that eq.(\ref{qnng-8}) corresponds to the following table integral for the Hermite polynomials~\cite[2.20.3.17]{prudnikov2002-2}:

\[
\int\limits_{ - \infty }^\infty  {e^{ - \left( {x - y} \right)^2 } H_n \left( y \right)dy}  = \sqrt \pi  \left( {2x} \right)^n .
\]

Trivial computations lead to its analytical expression (\ref{qnng-5}). As it is shown in the above-performed computations, its substitution at (\ref{hf-ogh}) leads to the exact expression of the Husimi function (\ref{hf-gh}).

One needs to note number of papers, which use both Wigner function (\ref{wf-gen}) and its simplest Gaussian smoothing function (\ref{hf-gen}) in the comparative studies of the various quantum systems and show advantage of each function for classical description of the certain quantum system of the particular behaviour~\cite{takahashi1985,takahashi1986,casida1987,harriman1988,morrison1991,harriman1993,nagiyev1998,hentschel2003,terraneo2005,hongyi2007,jafarov2007,andreev2011,berry2014,tsukiji2016}.

\section{Computation of the Husimi function of a semiconfined harmonic oscillator model}

Before starting the main computations, we want to provide brief information about the model of a semiconfined harmonic oscillator. As it was noted in the Introduction of the current paper, the model was introduced as an exactly-solvable generalization of the one-dimensional non-relativistic canonical quantum harmonic oscillator, where its constant effective mass was replaced by the effective mass changing by position~\cite{jafarov2021}. The model is semiconfined, i.e. its wavefunctions of the stationary states vanish at both values of the position $x=-a$ and $x \to +\infty$. Surprisingly, its energy spectrum completely overlaps with the energy spectrum of the standard non-relativistic canonical quantum harmonic oscillator, but, its wavefunctions of the stationary states are expressed through the generalized Laguerre polynomials as follows:

\be
\label{wf-sc-0}
\psi _n \left( x \right) \equiv \psi _n^{SC} \left( x \right) = C_n^{SC} \left( {1 + \frac{x}{a}} \right)^{\lambda _0 ^2 a^2 } e^{ - \lambda _0 ^2 a\left( {x + a} \right)} L_n^{\left( {2\lambda _0 ^2 a^2 } \right)} \left( {2\lambda _0 ^2 a\left( {x + a} \right)} \right),
\ee
where the normalization factor equals to

\be
\label{c-n}
C_n^{SC}  = \left( { - 1} \right)^n \left( {2\lambda _0 ^2 a^2 } \right)^{\lambda _0 ^2 a^2  + \frac{1}{2}} \sqrt {\frac{{n!}}{{a\Gamma \left( {n + 2\lambda _0 ^2 a^2  + 1} \right)}}} .
\ee

There are a lot of physical phenomena, which can be explained more precisely by applying the theoretical models of asymmetrical quantum systems. For example, the growth of sub-micron or nano-sized crystal structures on a substrate~\cite{saito1996,sunagawa2001,dost2007}, fabrication of advanced nano-scale QFET structures~\cite{sze2002,datta2016}, extensions of the various stochastic processes~\cite{schoutens2000} as well as a number of problems belonging to econophysics exhibiting initial restrictions of the infinitely high wall behavior~\cite{ahn2017} are just a few of possible examples. In general, asymmetrical physical structures belonging to material science are in fashion thanks to the progress in the advanced epitaxy technologies allowing the growth of alternating nano-sized layers of semiconductor materials. A semiconfined harmonic oscillator model having a position-dependent mass and exhibiting the energy spectrum completely overlapping with the energy spectrum of the one-dimensional non-relativistic canonical quantum harmonic oscillator is the best candidate for such an asymmetrical quantum system.

The model of the semiconfined harmonic oscillator with a mass changing with position~\cite{jafarov2021} also was generalized to the case of the suddenly applied external homogeneous field~\cite{jafarov2022}. Motivation for studying this generalized problem was based on assumption that an external field applied to the asymmetrical quantum system can adjust its asymmetry and this will lead to large enhancements of the second-order nonlinear optical effects in the structures described through such asymmetrical quantum models~\cite{weisbuch1991,zhang2003,zhang2003mplb,zhang2004}.

The following analytical expression of the semiconfined oscillator wavefunctions of the stationary states under the external field in terms of the generalized Laguerre polynomials have been obtained:

\be
\label{wf-gsc}
\psi _n \left( x \right) \equiv \psi _n^{gSC} \left( x \right) =  C_n^{gSC} \left( {1 + \frac{x}{a}} \right)^{\lambda _0 ^2 a^2 } e^{ - \lambda _0 ^2 ag_0 \left( {x + a} \right)} L_n^{\left( {2\lambda _0 ^2 a^2 } \right)} \left( {2\lambda _0 ^2 ag_0 \left( {x + a} \right)} \right),
\ee
where,

\be
\label{c-ng}
C_n^{gSC}=g_0 ^{\lambda _0 ^2 a^2  + \frac{1}{2}} C_n^{SC},
\ee
with the normalization factor $C_n^{SC}$ same as from (\ref{c-n}) and the parameter $g_0$ defined as

\be
\label{g0}
g_0  = \sqrt {1 + \frac{{2g}}{{m_0 \omega^2 a}}} .
\ee

One can easily observe that when $g=0$, i.e. an external homogeneous field simply disappears, then the parameter $g_0$ is equal to one, and the wavefunction (\ref{wf-gsc}) reduces to the wavefunction (\ref{wf-sc-0}). Therefore, we can compute the smoothed Wigner function of the semiconfined quantum harmonic oscillator under the action of the homogeneous external field by substitution of its wavefunctions of the stationary states (\ref{wf-gsc}) at (\ref{hf-gen}) and then to explore special case $g=0$ of the computed expression of the distribution function that will correspond to the smoothed Wigner function of the semiconfined quantum harmonic oscillator described by the wavefunctions (\ref{wf-sc-0}).

Here the question can arise that why we decided to compute the Gaussian smoothed Wigner function or Husimi function, but not the Wigner function itself? The main problem is not related to physics, but to mathematics. Starting the computation of the analytical expression by the canonical method introduced in~\cite{wigner1932}, one could easily observe that the integrand from eq.(\ref{wf-gen}) simply diverges and this fact makes it impossible to perform further calculations. However, Gaussian smoothing of eq.(\ref{wf-gen}) restricts this divergence and allows to perform the presented below computations (cf. with~\cite{jafarov2008}, which succeeds with computation of the exact expression of the Wigner function of the one-dimensional parabose oscillator, but not its Husimi function, due to that the momentum and position operators commute in a non-canonical manner). Then, taking into account that the Husimi function of the quantum system can be as informative as the Wigner function itself, we decided to go on with the computation of the Gaussian smoothed version.

We start our computation of the Husimi function from its following definition:

\be
\label{hf-gsc1}
\bar W_n^g \left( {p,x} \right) = \frac{{\lambda _0 }}{{2\pi \hbar \sqrt \pi  }}\left| {\bar Q_n^g } \right|^2  = \frac{1}{{2\pi \hbar C_{N0} ^2 }}\left| {\bar Q_n^g } \right|^2 ,
\ee
where,

\be
\label{qng-1}
\bar Q_n^g  = \int\limits_{ - a}^\infty  {\psi _n^{gSC} \left( {x'} \right)e^{ - i\frac{{px'}}{\hbar } - \frac{{\lambda _0 ^2 }}{2}\left( {x - x'} \right)^2 } dx'} .
\ee

The main difference of (\ref{qng-1}) from (\ref{qnng-1}) is that now it is necessary to compute an integral with lower boundary $-a$. Here, we will slightly change the methods used in the previous section, i.e. we will start our computations from the ground state $n=0$. Then one has

\be
\label{hf0-gsc1}
\bar W_0^g \left( {p,x} \right) = \frac{{\lambda _0 }}{{2\pi \hbar \sqrt \pi  }}\left| {\bar Q_0^g } \right|^2  ,
\ee
where,

\be
\label{q0g-1}
\bar Q_0^g  = \int\limits_{ - a}^\infty  {\psi _0^{gSC} \left( {x'} \right)e^{ - i\frac{{px'}}{\hbar } - \frac{{\lambda _0 ^2 }}{2}\left( {x - x'} \right)^2 } dx'} .
\ee

Substitution of the ground-state wavefunction

\be
\label{psi0-sc}
\psi _0^{gSC} \left( x \right) = C_0^{gSC} \left( {1 + \frac{x}{a}} \right)^{\lambda _0 ^2 a^2 } e^{ - \lambda _0 ^2 a^2 g_0 \left( {1 + \frac{x}{a}} \right)} ,\quad C_0^{gSC} = \frac{{\left( {2\lambda _0 ^2 a^2 g_0 } \right)^{\lambda _0 ^2 a^2  + \frac{1}{2}} }}{{\sqrt {a\Gamma \left( {2\lambda _0 ^2 a^2  + 1} \right)} }},
\ee
at (\ref{q0g-1}) yields

\be
\label{q0g-2}
\bar Q_0^g  = C_0^{gSC} \int\limits_{ - a}^\infty  {\left( {1 + \frac{{x'}}{a}} \right)^{b^2 } e^{ - i\frac{{px'}}{\hbar } - \frac{{\lambda _0 ^2 }}{2}\left( {x - x'} \right)^2  - b^2 g_0 \left( {1 + \frac{{x'}}{a}} \right)} dx'} ,
\ee
where, $b=\lambda _0 a$. Introduction of new dimensionless variable $\bar y = 1 + \frac{{x'}}{a}$ allows to shift the lower boundary of the integral from $-a$ to zero as follows:

\be
\label{q0g-3}
\bar Q_0^g  = aC_0^{gSC} \int\limits_0^\infty  {\bar y^{b^2 } e^{ - i\frac{{pa}}{\hbar }\left( {\bar y - 1} \right) - \frac{{\lambda _0 ^2 }}{2}\left( {x - a\left( {\bar y - 1} \right)} \right)^2  - b^2 g_0 \bar y} d\bar y} .
\ee

The exponential function of the integrand can be simplified by introducing the following notations:

\[
\bar \beta _1  =  - bz,\quad z = z_1  + iz_2 ,\quad z_1  = bg_0  - b_1 ,\quad z_2  = \eta ,\quad b_1  = \xi  + b,\quad \bar \beta _0  =  - \frac{1}{2}b_1 ^2  + ib\eta .
\]

One obtains:

\be
\label{q0g-4}
\bar Q_0^{g}  = aC_0^{gSC} \int\limits_0^\infty  {\bar y^{b^2 } e^{ - \frac{1}{2}b^2 \bar y^2  + \bar \beta _1 \bar y + \bar \beta _0 } d\bar y} .
\ee

This integral can be computed exactly by employing here specific Maclaurin series expansion of the exponential function. It is a table integral and can be found from \cite[2.3.15.3]{prudnikov2002} as follows:

\be
\label{t-i}
\int\limits_0^\infty  {y^{\alpha  - 1} e^{ - py^2  - qy} dy}  = \Gamma \left( \alpha  \right)\left( {2p} \right)^{ - \alpha /2} e^{\frac{{q^2 }}{{8p}}} D_{ - \alpha } \left( {\frac{q}{{\sqrt {2p} }}} \right),\quad {\mathop{\rm Re}\nolimits} \alpha ,{\mathop{\rm Re}\nolimits} p > 0,
\ee
where, $D_{\nu } \left( z \right)$ is a parabolic cylinder function of the following definition in terms of the $_1 F_1$ hypergeometric functions:

\be
\label{pcf}
D_\nu  \left( z \right) = \sqrt \pi  2^{\frac{\nu }{2}} e^{ - \frac{{z^2 }}{4}} \left[ {\frac{1}{{\Gamma \left( {\frac{{1 - \nu }}{2}} \right)}}\,_1 F_1 \left( {\begin{array}{*{20}c}
   { - \frac{\nu }{2}}  \\
   {\frac{1}{2}}  \\
\end{array};\frac{{z^2 }}{2}} \right) - \frac{{\sqrt 2 z}}{{\Gamma \left( { - \frac{\nu }{2}} \right)}}\,_1 F_1 \left( {\begin{array}{*{20}c}
   {\frac{{1 - \nu }}{2}}  \\
   {\frac{3}{2}}  \\
\end{array};\frac{{z^2 }}{2}} \right)} \right].
\ee

Substitution of (\ref{t-i}) at (\ref{q0g-4}) and taking into account analytical expression of $C_0^{gSC}$ from (\ref{psi0-sc}) yields

\be
\label{q0g-5}
\bar Q_0^{g}  = \frac{{\left( {2bg_0 } \right)^{b^2  + \frac{1}{2}}  }}{{\sqrt {\lambda _0 \Gamma \left( {2b^2  + 1} \right)} }}\Gamma \left( {b^2  + 1} \right)e^{\frac{{z^2 }}{4} + \bar \beta _0 } D_{ - b^2  - 1} \left( z \right).
\ee

Finally, substitution of (\ref{q0g-5}) at (\ref{hf0-gsc1}) yields the Husimi function of the ground state:

\begin{eqnarray}
\label{hf-0}
 &\bar W_0^{g} \left( {p,x} \right) = \frac{1}{{\pi \hbar }} \left( {g_0 \lambda _0 a} \right)^{2\lambda _0 ^2 a^2  + 1} \frac{{\Gamma \left( {\lambda _0 ^2 a^2  + 1} \right)}}{{\Gamma \left( {\lambda _0 ^2 a^2  + \frac{1}{2}} \right)}} e^{ - \frac{1}{{\hbar \omega }}\left( {\frac{{p^2 }}{{2m_0 }} + \frac{{m_0 \omega ^2 }}{2}\left( {x + a\left( {g_0  + 1} \right)} \right)^2  - m_0 \omega ^2 a^2 g_0 ^2 } \right)}\\ 
 & \times  D_{ - \left( {\lambda _0 ^2 a^2   + 1} \right)} \left( { - \lambda _0 \left( {x + a\left( {1 - g_0 } \right) - i\frac{p}{{m_0 \omega }}} \right)} \right)D_{ - \left( {\lambda _0 ^2 a^2   + 1} \right)} \left( { - \lambda _0 \left( {x + a\left( {1 - g_0 } \right) + i\frac{p}{{m_0 \omega }}} \right)} \right). \nonumber
\end{eqnarray}

Absence of the external field corresponding to the case $g=0$ ($g_0=1$) reduces this expression to the following analytical expression of the Husimi function of the semiconfined quantum harmonic oscillator ground state:

\begin{eqnarray}
\label{hf-00}
 &\bar W_0^{0} \left( {p,x} \right) = \frac{1}{{\pi \hbar }} \left( {\lambda _0 a} \right)^{2\lambda _0 ^2 a^2  + 1} \frac{{\Gamma \left( {\lambda _0 ^2 a^2  + 1} \right)}}{{\Gamma \left( {\lambda _0 ^2 a^2  + \frac{1}{2}} \right)}}e^{ - \frac{1}{{\hbar \omega }}\left( {\frac{{p^2 }}{{2m_0 }} + \frac{{m_0 \omega ^2 }}{2}\left( {x^2  + 4ax + 2a^2 } \right)} \right)} \\ 
  &\times D_{ - \left( {\lambda _0 ^2 a^2   + 1} \right)} \left( { - \lambda _0 \left( {x - i\frac{p}{{m_0 \omega }}} \right)} \right)D_{ - \left( {\lambda _0 ^2 a^2   + 1} \right)} \left( { - \lambda _0 \left( {x + i\frac{p}{{m_0 \omega }}} \right)} \right) . \nonumber
 \end{eqnarray}

Now, we can extend these computations to arbitrary states. One needs to substitute analytical expression of the wavefunction of the arbitrary states (\ref{wf-gsc}) at (\ref{qng-1}):

\be
\label{qng-2}
\bar Q_n^g  = C_n^{gSC} \int\limits_{ - a}^\infty  {\left( {1 + \frac{{x'}}{a}} \right)^{b^2 } e^{ - \frac{{\lambda _0 ^2 }}{2}\left( {x - x'} \right)^2 - b^2 g_0 \left( {1 + \frac{{x'}}{a}} \right) - i\frac{{px'}}{\hbar } } L_n^{\left( {2b^2 } \right)} \left( {2b^2 g_0 \left( {1 + \frac{{x'}}{a}} \right)} \right)dx'} .
\ee

We continue with the operator method and apply the mathematical trick for the Laguerre polynomials similar to the trick, which was used in the case of the Hermite polynomials (\ref{hp-1}):

\be
\label{lp-1}
L_n^{\left( {2b^2 } \right)} \left( {2b^2 g_0 \left( {1 + \frac{{x'}}{a}} \right)} \right) \cdot e^{ - i\frac{{px'}}{\hbar }}  = L_n^{\left( {2b^2 } \right)} \left( {2i\lambda _0 bg_0 \hbar \frac{d}{{dp}} + 2b^2 g_0 } \right) \cdot e^{ - i\frac{{px'}}{\hbar }} .
\ee

Its substitution at (\ref{qng-2}) yields

\be
\label{qng-3}
\bar Q_n^g  = C_n^{gSC} L_n^{\left( {2b^2 } \right)} \left( {2i\lambda _0 bg_0 \hbar \frac{d}{{dp}} + 2b^2 g_0 } \right) \cdot \int\limits_{ - a}^\infty  {\left( {1 + \frac{{x'}}{a}} \right)^{b^2 } e^{ - \frac{{\lambda _0 ^2 }}{2}\left( {x - x'} \right)^2 - b^2 g_0 \left( {1 + \frac{{x'}}{a}} \right) - i\frac{{px'}}{\hbar }  } dx'} .
\ee

From this definition, one can easily write down the following connection between $\bar Q_n^g$ and $\bar Q_0^g$:

\be
\label{qng-4}
\bar Q_n^g  = \frac{{C_n^{gSC} }}{{C_0^{gSC} }}L_n^{\left( {2b^2 } \right)} \left( {2i\lambda _0 bg_0 \hbar \frac{d}{{dp}} + 2b^2 g_0 } \right) \cdot \bar Q_0^g .
\ee

Taking into account that

\[
\frac{{C_n^{gSC} }}{{C_0^{gSC} }} = \left( { - 1} \right)^n \sqrt {\frac{{n!}}{{\left( {2b^2  + 1} \right)_n }}} ,
\]
one obtains

\be
\label{qng-5}
\bar Q_n^g  = \left( { - 1} \right)^n \sqrt {\frac{{n!}}{{\left( {2b^2  + 1} \right)_n }}} L_n^{\left( {2b^2 } \right)} \left( {2i\lambda _0 bg_0 \hbar \frac{d}{{dp}} + 2b^2 g_0 } \right) \cdot \bar Q_0^g .
\ee

Its substitution at (\ref{hf-gsc1}) yields

\be
\label{hf-op0}
\bar W_n^g \left( {p,x} \right) = \frac{{\lambda _0 }}{{2\pi \hbar \sqrt \pi  }} \frac{{n!}}{{\left( {2\lambda _0 ^2 a^2  + 1} \right)_n }}\left|L_n^{\left( {2b^2 } \right)} \left( {2b^2 g_0 + 2i\lambda _0 bg_0 \hbar \frac{d}{{dp}} } \right) \cdot \bar Q_0^g\right|^2.
\ee

This is an analytical expression of the Husimi function of the arbitrary $n$ state in the form of the differential operator acting to the Husimi function of the ground state (\ref{hf-0}). In case of the absence of the external homogeneous field $g=0$ ($g_0=1$) it reduces to the following expression of the Husimi function of the arbitrary $n$ state:

\be
\label{hf-opg0}
\bar W_n^0 \left( {p,x} \right) = \frac{{\lambda _0 }}{{2\pi \hbar \sqrt \pi  }} \frac{{n!}}{{\left( {2\lambda _0 ^2 a^2  + 1} \right)_n }}\left|L_n^{\left( {2b^2 } \right)} \left( {2b^2 + 2i\lambda _0 b \hbar \frac{d}{{dp}} } \right) \cdot \bar Q_0^0\right|^2.
\ee

Here, $\bar Q_0^0$ is the special case of $\bar Q_0^g$ defined via (\ref{q0g-5}) when $g=0$ ($g_0=1$).

One can continue our computations and extract from (\ref{hf-op0}) an analytical expression of the Husimi function that will generalize eq.(\ref{hf-gh}). To achieve this, one needs to use the series expansion of $_1F_1$ hypergeometric functions from the definition of a parabolic cylinder function (\ref{pcf}) and then to apply the following generating function for the generalized Laguerre polynomials to definition (\ref{hf-0}) of $\bar W_0^g \left( {p,x} \right)$:

\[
\sum\limits_{n = 0}^\infty  {L_n^{\left( \alpha  \right)} \left( x \right)t^n }  = \left( {1 - t} \right)^{ - \alpha  - 1} e^{\frac{{xt}}{{t - 1}}} .
\]

However, we will not do it, because the Husimi function (\ref{hf-op0}) is a completely elegant operator expression and quite useful for further discussions of the phase-space properties of the quantum system under consideration. Instead of it, we will go on with the second method, which is the direct computation of the Husimi function.

We generalize the computations of (\ref{q0g-2}) to case of $\bar Q_n^g$ defined through (\ref{qng-2}). Introduction of new dimensionless variable $\bar y = 1 + \frac{{x'}}{a}$ again shifts the lower boundary of the integral from $-a$ to zero as follows:

\be
\label{qng-d1}
\bar Q_n^g  = aC_n^{gSC} \int\limits_0^\infty  {\bar y^{b^2 } e^{ - i\frac{{pa}}{\hbar }\left( {\bar y - 1} \right) - \frac{{\lambda _0 ^2 }}{2}\left( {x - a\left( {\bar y - 1} \right)} \right)^2  - b^2 g_0 \bar y} L_n^{\left( {2b^2 } \right)} \left( {2b^2 g_0 \bar y} \right)d\bar y} .
\ee

Let's use the following finite sum expansion of the generalized Laguerre polynomials

\[
L_n^{\left( {2b^2 } \right)} \left( {2b^2 g_0 \bar y} \right) = \frac{{\left( {2b^2  + 1} \right)_n }}{{n!}}\sum\limits_{k = 0}^n {\frac{{\left( { - n} \right)_k }}{{\left( {2b^2  + 1} \right)_k }}\frac{{\left( {2b^2 g_0 } \right)^k }}{{k!}}{\bar y}^k } ,
\]
and substitute it at (\ref{qng-d1}). Then, we have

\be
\label{qng-d2}
\bar Q_n^g  = aC_n^{gSC} \frac{{\left( {2b^2  + 1} \right)_n }}{{n!}}\sum\limits_{k = 0}^n {\frac{{\left( { - n} \right)_k }}{{\left( {2b^2  + 1} \right)_k }}\frac{{\left( {2b^2 g_0 } \right)^k }}{{k!}}\int\limits_0^\infty  {\bar y^{b^2  + k} e^{ - i\frac{{pa}}{\hbar }\left( {\bar y - 1} \right) - \frac{{\lambda _0 ^2 }}{2}\left( {x - a\left( {\bar y - 1} \right)} \right)^2  - b^2 g_0 \bar y} d\bar y} } .
\ee

The exponential function of the integrand again can be simplified allowing us to rewrite eq.(\ref{qng-d2}) as follows:

\be
\label{qng-d3}
\bar Q_n^g  = aC_n^{gSC} \frac{{\left( {2b^2  + 1} \right)_n }}{{n!}}\sum\limits_{k = 0}^n {\frac{{\left( { - n} \right)_k }}{{\left( {2b^2  + 1} \right)_k }}\frac{{\left( {2b^2 g_0 } \right)^k }}{{k!}}\int\limits_0^\infty  {\bar y^{b^2  + k} e^{ - \frac{1}{2}b^2 \bar y^2  + \bar \beta _1 \bar y + \bar \beta _0 } d\bar y} } .
\ee

This integral again leads to parabolic cylinder function due to (\ref{t-i}). Some trivial simplifications yield:

\be
\label{qng-d4}
\bar Q_n^g  = \left( { - 1} \right)^n \frac{{\left( {2bg_0 } \right)^{b^2  + \frac{1}{2}} }}{{\sqrt {\lambda _0 \Gamma \left( {2b^2  + 1} \right)} }}e^{\frac{{z^2 }}{4} + \bar \beta _0 } \sqrt {\frac{{\left( {2b^2  + 1} \right)_n }}{{n!}}} \sum\limits_{k = 0}^n {\frac{{\left( { - n} \right)_k }}{{\left( {2b^2  + 1} \right)_k }}\frac{{\left( {2bg_0 } \right)^k }}{{k!}}\Gamma \left( {b^2  + k + 1} \right)D_{ - \left( {b^2  + k + 1} \right)} \left( z \right)} .
\ee

Substitution of (\ref{qng-4}) at (\ref{hf-gsc1}) yields the analytical expression of the Husimi function of the semiconfined quantum harmonic oscillator described by the wavefunctions (\ref{wf-gsc}):

\begin{eqnarray}
\label{hf-4}
 &\bar W_n^{g} \left( {p,x} \right) = \frac{1}{{\pi \hbar }}e^{ - \frac{1}{{\hbar \omega }}\left( {\frac{{p^2 }}{{2m_0 }} + \frac{{m_0 \omega ^2 }}{2}\left( {x + a\left( {g_0  + 1} \right)} \right)^2  - m_0 \omega ^2 a^2 g_0 ^2 } \right)} \left( {g_0 \lambda _0 a} \right)^{2\lambda _0 ^2 a^2  + 1} \frac{{\Gamma \left( {\lambda _0 ^2 a^2  + 1} \right)}}{{\Gamma \left( {\lambda _0 ^2 a^2  + \frac{1}{2}} \right)}}\frac{{\left( {2\lambda _0 ^2 a^2  + 1} \right)_n }}{{n!}} \\ 
 & \times  \scriptstyle{\sum\limits_{k,s = 0}^n {\frac{{\left( { - n} \right)_k \left( { - n} \right)_s \left( {\lambda _0 ^2 a^2  + 1} \right)_k \left( {\lambda _0 ^2 a^2  + 1} \right)_s }}{{\left( {2\lambda _0 ^2 a^2  + 1} \right)_k \left( {2\lambda _0 ^2 a^2  + 1} \right)_s }}\frac{{\left( {2g_0 \lambda _0 a} \right)^{k + s} }}{{k!s!}}D_{ - \left( {\lambda _0 ^2 a^2  + k + 1} \right)} \left( { - \lambda _0 \left( {x + a\left( {1 - g_0 } \right) - i\frac{p}{{m_0 \omega }}} \right)} \right)D_{ - \left( {\lambda _0 ^2 a^2  + s + 1} \right)} \left( { - \lambda _0 \left( {x + a\left( {1 - g_0 } \right) + i\frac{p}{{m_0 \omega }}} \right)} \right)}}. \nonumber
\end{eqnarray}

Absence of the external field corresponding to the case $g=0$ reduces this expression to the following analytical expression of the Husimi function of the semiconfined quantum harmonic oscillator described by the wavefunctions (\ref{wf-sc-0}):

\begin{eqnarray}
\label{hf-g0}
 &\bar W_n^{0} \left( {p,x} \right) = \frac{1}{{\pi \hbar }}e^{ - \frac{1}{{\hbar \omega }}\left( {\frac{{p^2 }}{{2m_0 }} + \frac{{m_0 \omega ^2 }}{2}\left( {x^2  + 4ax + 2a^2 } \right)} \right)} \left( {\lambda _0 a} \right)^{2\lambda _0 ^2 a^2  + 1} \frac{{\Gamma \left( {\lambda _0 ^2 a^2  + 1} \right)}}{{\Gamma \left( {\lambda _0 ^2 a^2  + \frac{1}{2}} \right)}}\frac{{\left( {2\lambda _0 ^2 a^2  + 1} \right)_n }}{{n!}} \\ 
  &\times \sum\limits_{k,s = 0}^n {\frac{{\left( { - n} \right)_k \left( { - n} \right)_s \left( {\lambda _0 ^2 a^2  + 1} \right)_k \left( {\lambda _0 ^2 a^2  + 1} \right)_s }}{{\left( {2\lambda _0 ^2 a^2  + 1} \right)_k \left( {2\lambda _0 ^2 a^2  + 1} \right)_s }}\frac{{\left( {2\lambda _0 a} \right)^{k + s} }}{{k!s!}}D_{ - \left( {\lambda _0 ^2 a^2  + k + 1} \right)} \left( { - \lambda _0 \left( {x - i\frac{p}{{m_0 \omega }}} \right)} \right)D_{ - \left( {\lambda _0 ^2 a^2  + s + 1} \right)} \left( { - \lambda _0 \left( {x + i\frac{p}{{m_0 \omega }}} \right)} \right)} . \nonumber
 \end{eqnarray}

Our main goal is achieved, i.e. we succeeded with the computation of the exact expression of the Husimi function of the semiconfined quantum harmonic oscillator under the action of the external homogeneous field. In the final section, we are going to discuss the main properties of the model under study in the phase space.

\section{Discussions}

\begin{figure}[t!]
\begin{center}
\resizebox{0.40\textwidth}{!}{%
  \includegraphics{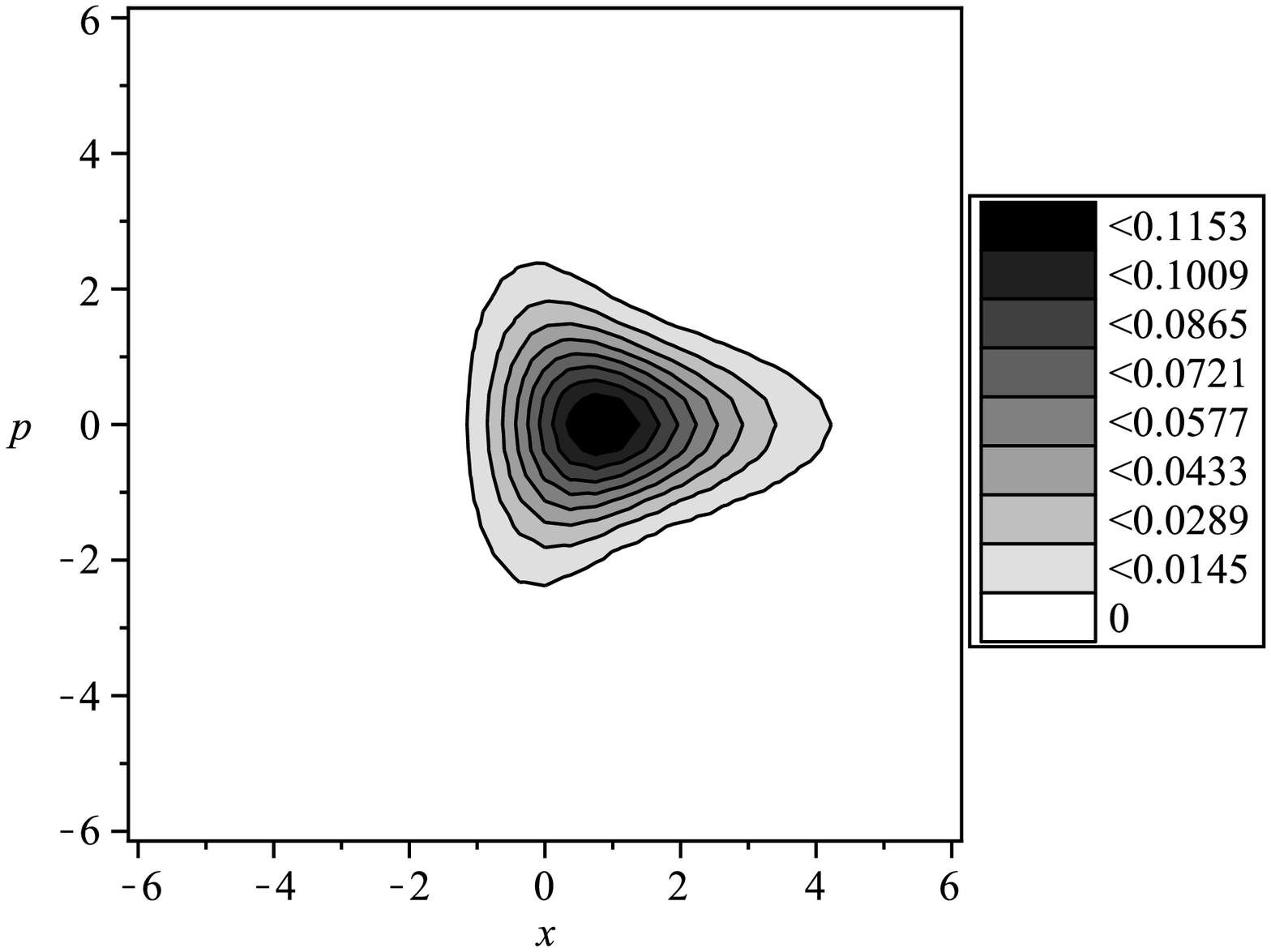}
}
\resizebox{0.40\textwidth}{!}{%
  \includegraphics{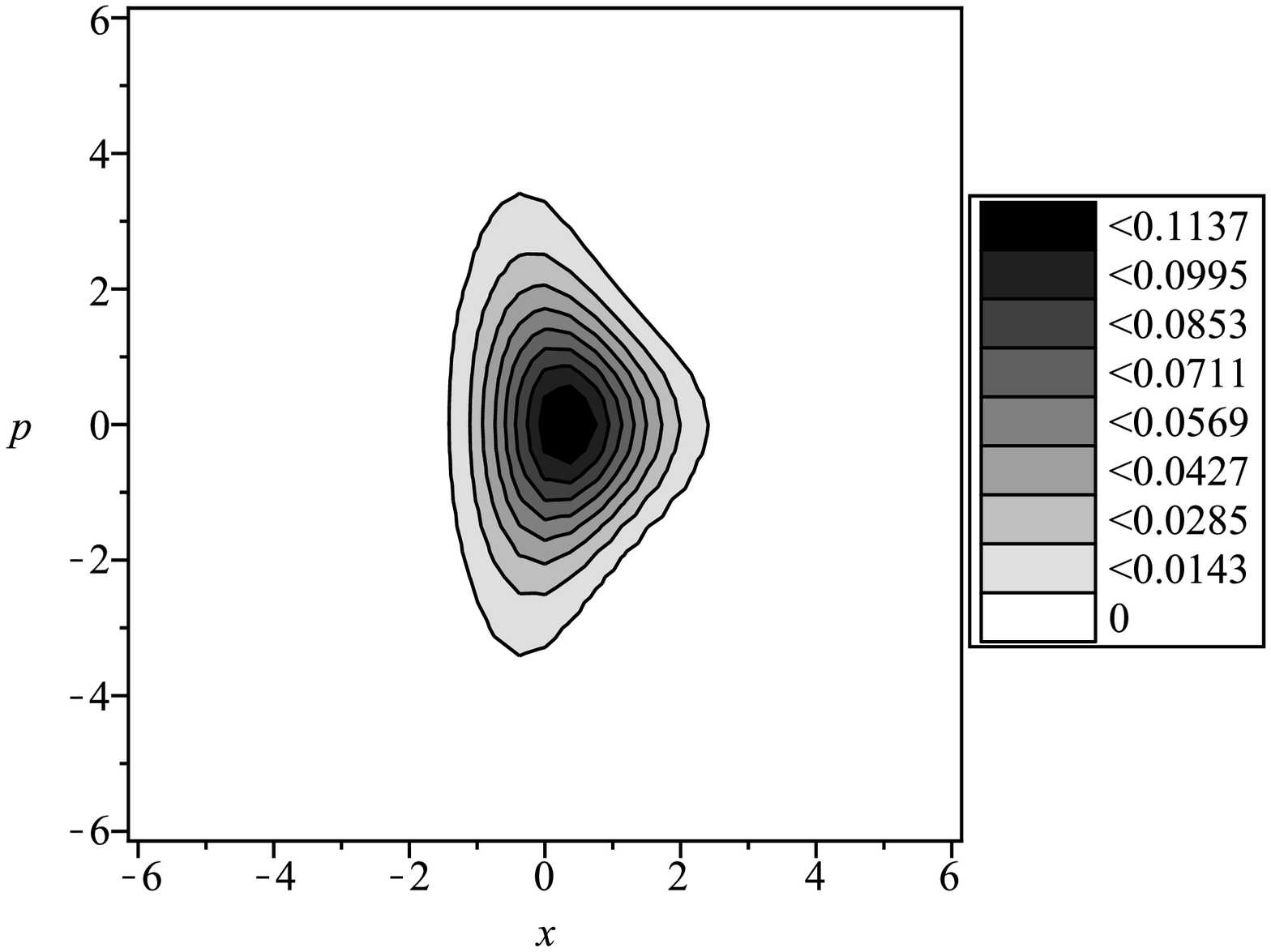}
}\\
\resizebox{0.40\textwidth}{!}{%
  \includegraphics{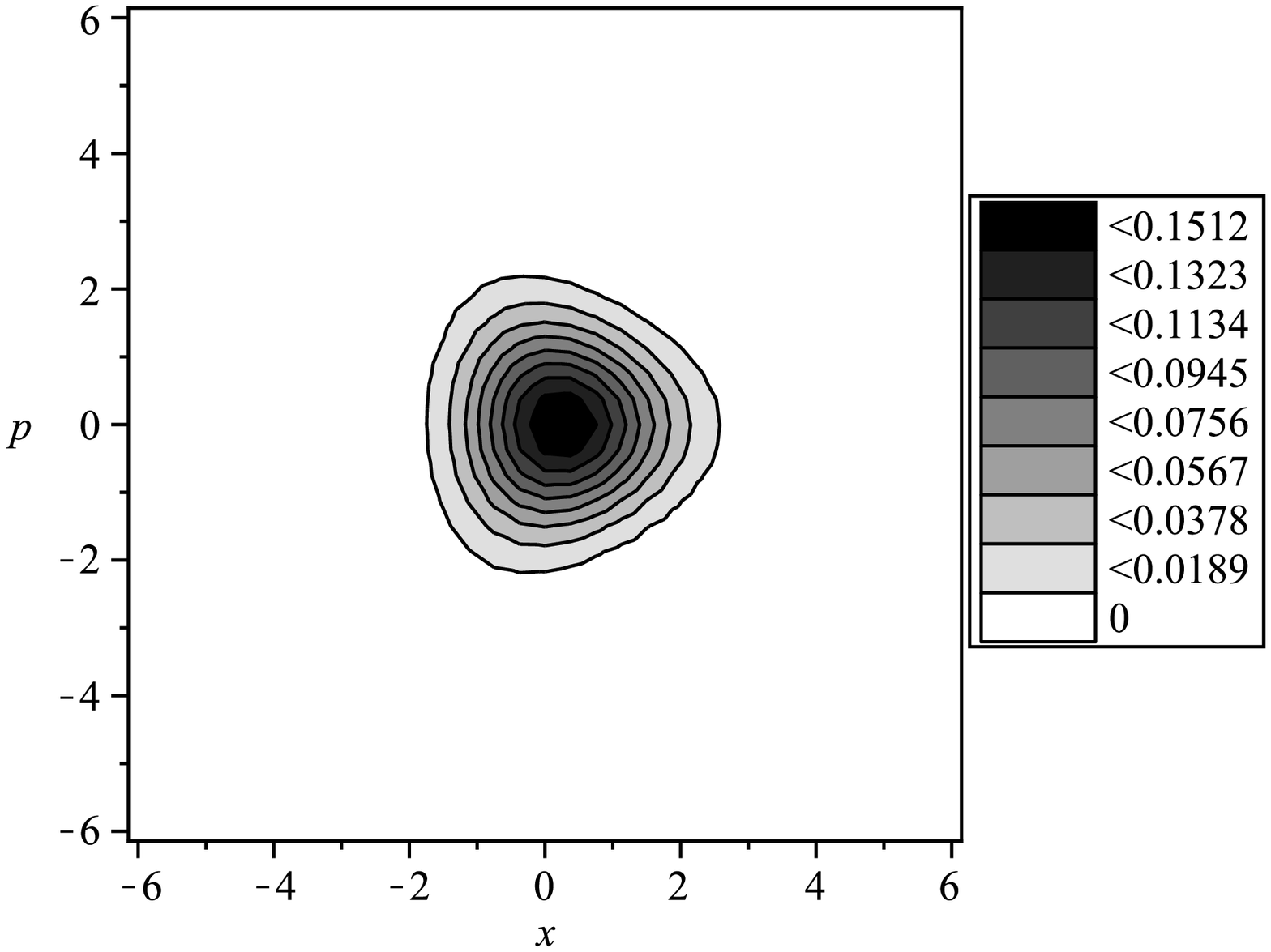}
}
\resizebox{0.40\textwidth}{!}{%
  \includegraphics{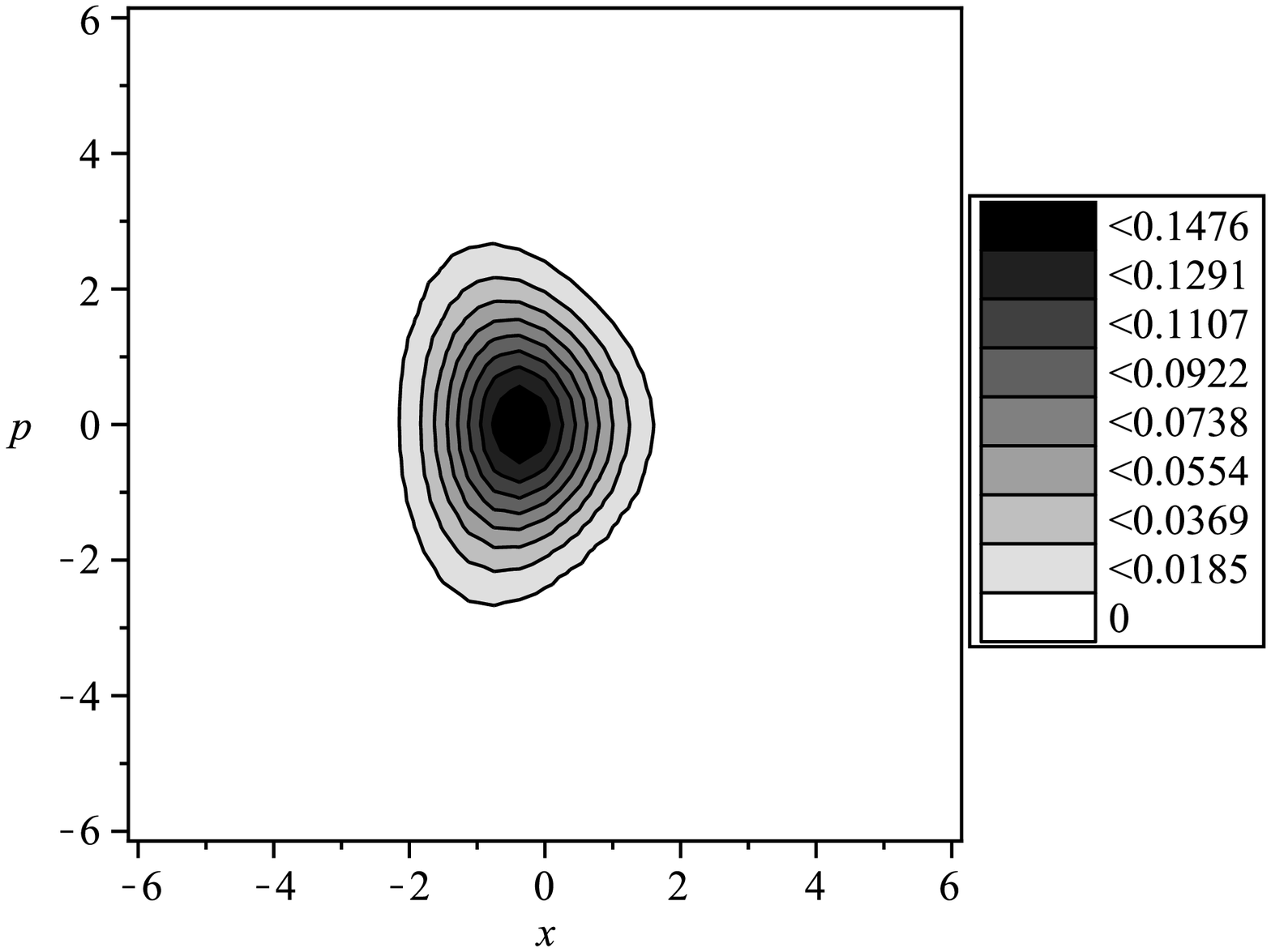}
}\\
\resizebox{0.40\textwidth}{!}{%
  \includegraphics{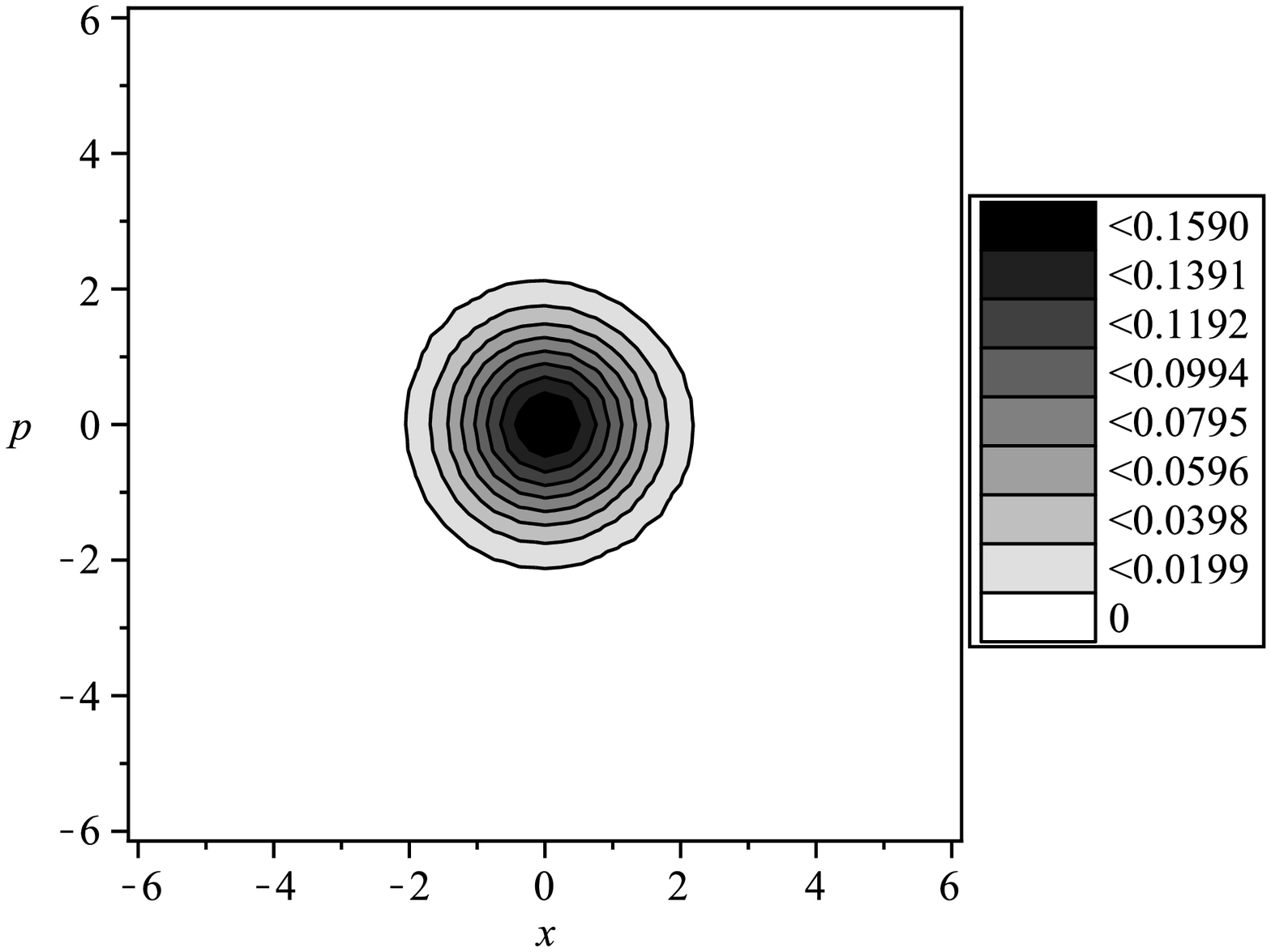}
}
\resizebox{0.40\textwidth}{!}{%
  \includegraphics{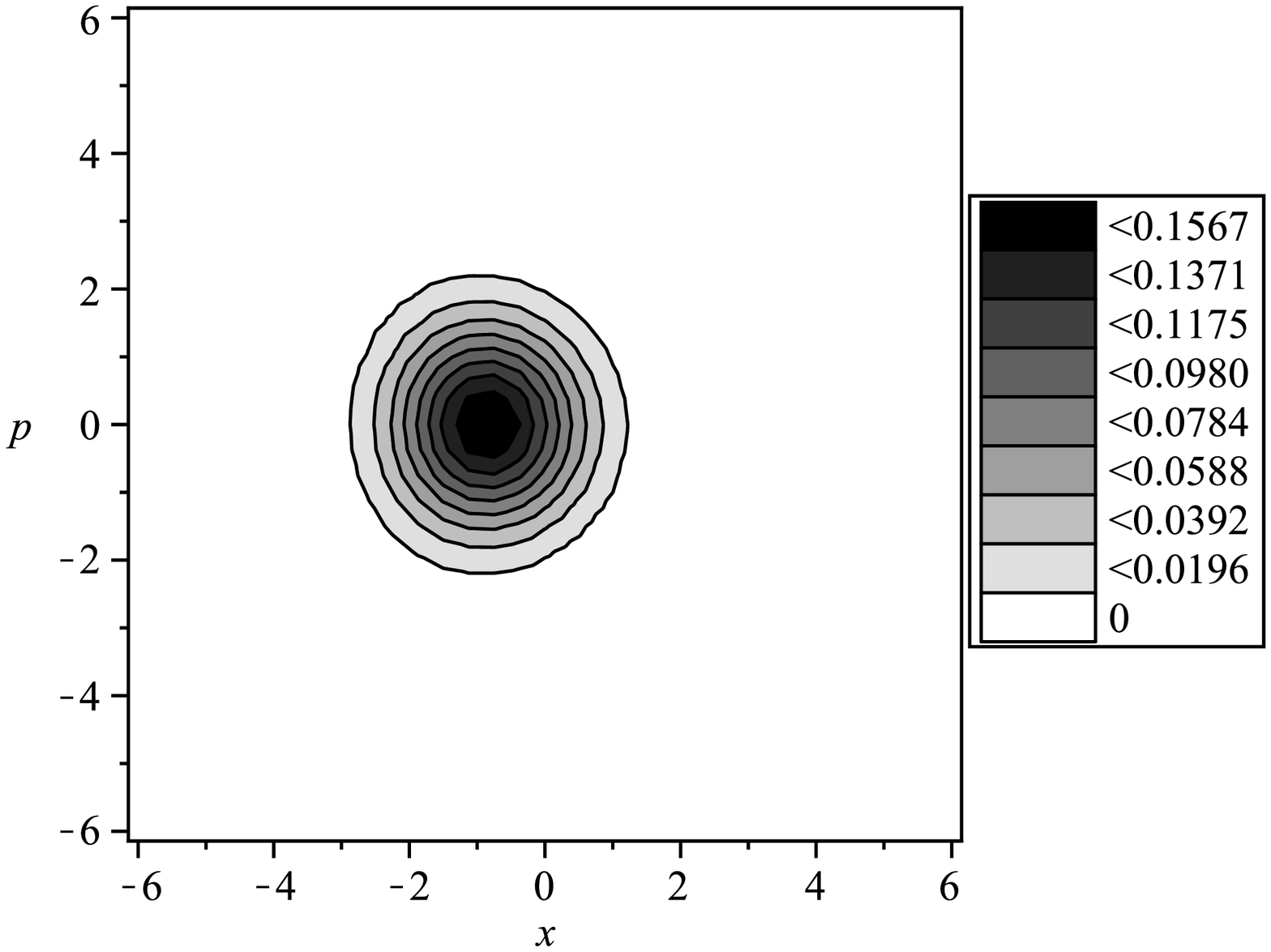}
}
\end{center}
\caption{Comparative plot of the semiconfined quantum harmonic oscillator Husimi function~(\ref{hf-4}) of the ground state ($n=0$) without an external field ($g=0$, left plots) and with an external field ($g=1$, right plots). Upper plots correspond to the confinement parameter value $a=0.5$, whereas, middle and lower plots correspond to the confinement parameter values $a=2$ and $a=12$ ($m_0=\omega=\hbar=1$).} 
\label{fig.1}
\end{figure}

In the preceding section, we were able to compute the exact expression of the Husimi function of the semiconfined quantum harmonic oscillator described by the wavefunctions (\ref{wf-sc-0}). Now, we are going to observe, what is the behavior of the semiconfined quantum system under study -- the main differences, which are hidden in the configuration space, but exhibited in the phase space. Therefore, we depict obtained analytical expression (\ref{hf-4}) of the semiconfined quantum harmonic oscillator Husimi function of the ground and first excited states without and with an external field. These plots are presented in figs. 1\&2.

We have to remind one of the main properties of the semiconfined quantum harmonic oscillator model \cite{jafarov2021,jafarov2022} that is exhibited in the $x$-configuration space: it was observed from the distribution of the probability densities $\left| {\psi _n^{SC} (x)} \right|^2$ and $\left| {\psi _n^{gSC} (x)} \right|^2$ computed from the wavefunctions (\ref{wf-sc-0}) and (\ref{wf-gsc}) that when the semiconfinement parameter $a$ is close to zero, then the quantum system under study stays close to the infinitely high wall and as the value of the parameter $a$ increased, the effect of the semiconfinement gradually disappeared and the behavior became harmonic oscillator-like. The phase space of the model's ground state in terms of the Husimi function perfectly overlaps with its above-highlighted behavior in the $x$-configuration space, as seen in fig.1. If no external homogeneous field is applied yet (left plots), the joint distribution of momentum and position shows that the maximum probability value is around zero. Any external homogeneous field $g \ne 0$ (right plots) then shifts and as well as smooths such a distribution. 

\begin{figure}[t!]
\begin{center}
\resizebox{0.40\textwidth}{!}{%
  \includegraphics{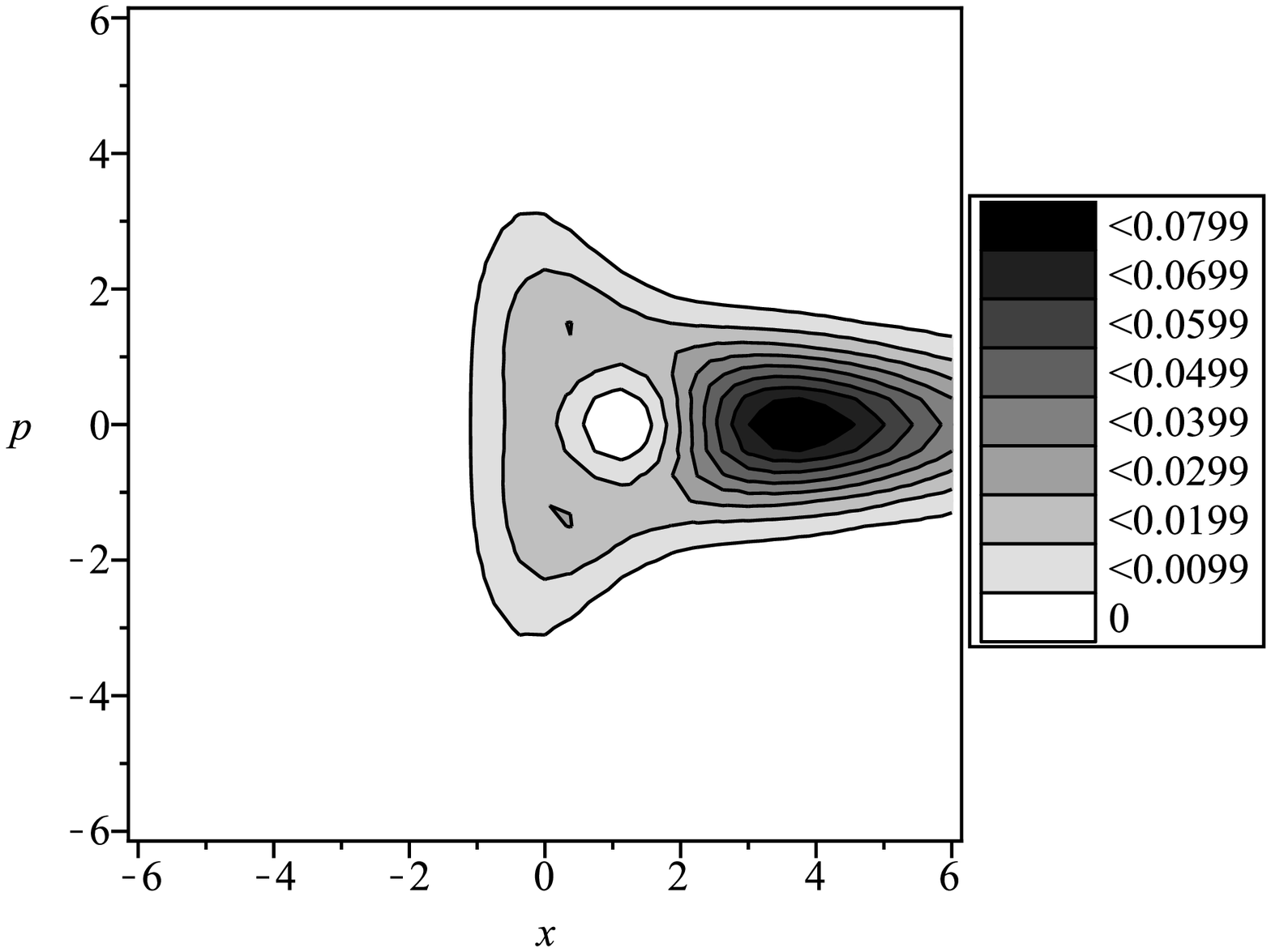}
}
\resizebox{0.40\textwidth}{!}{%
  \includegraphics{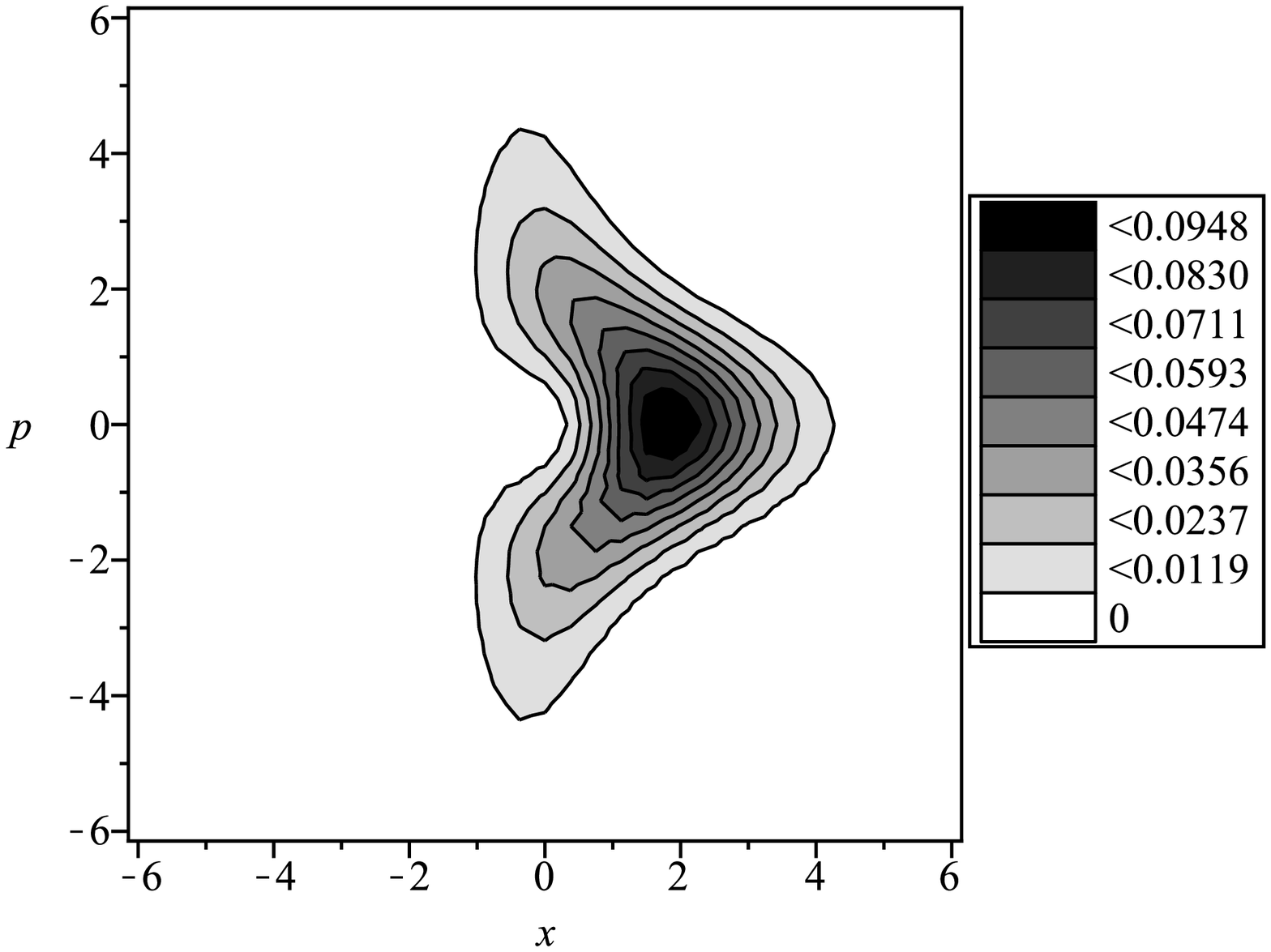}
}\\
\resizebox{0.40\textwidth}{!}{%
  \includegraphics{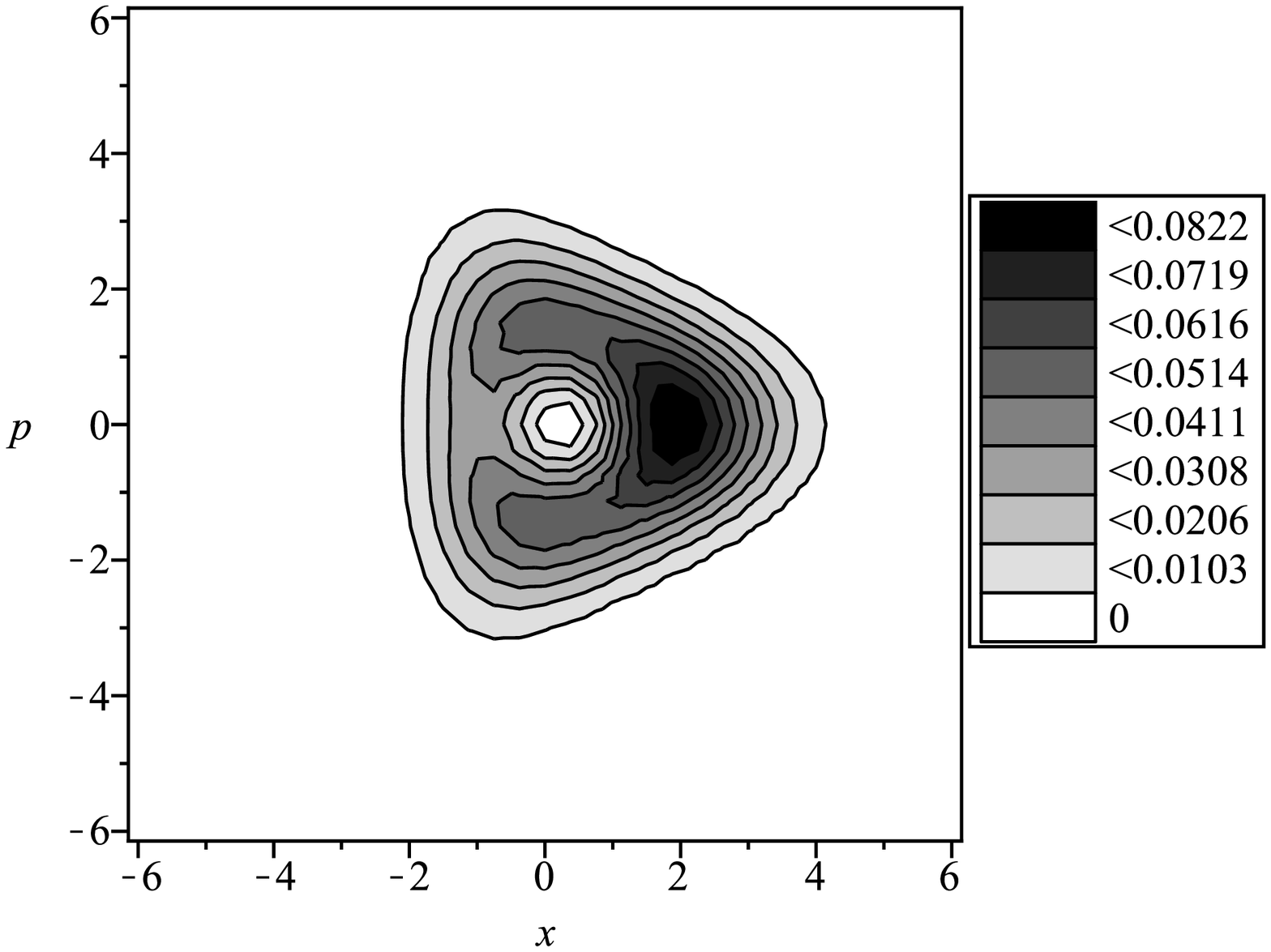}
}
\resizebox{0.40\textwidth}{!}{%
  \includegraphics{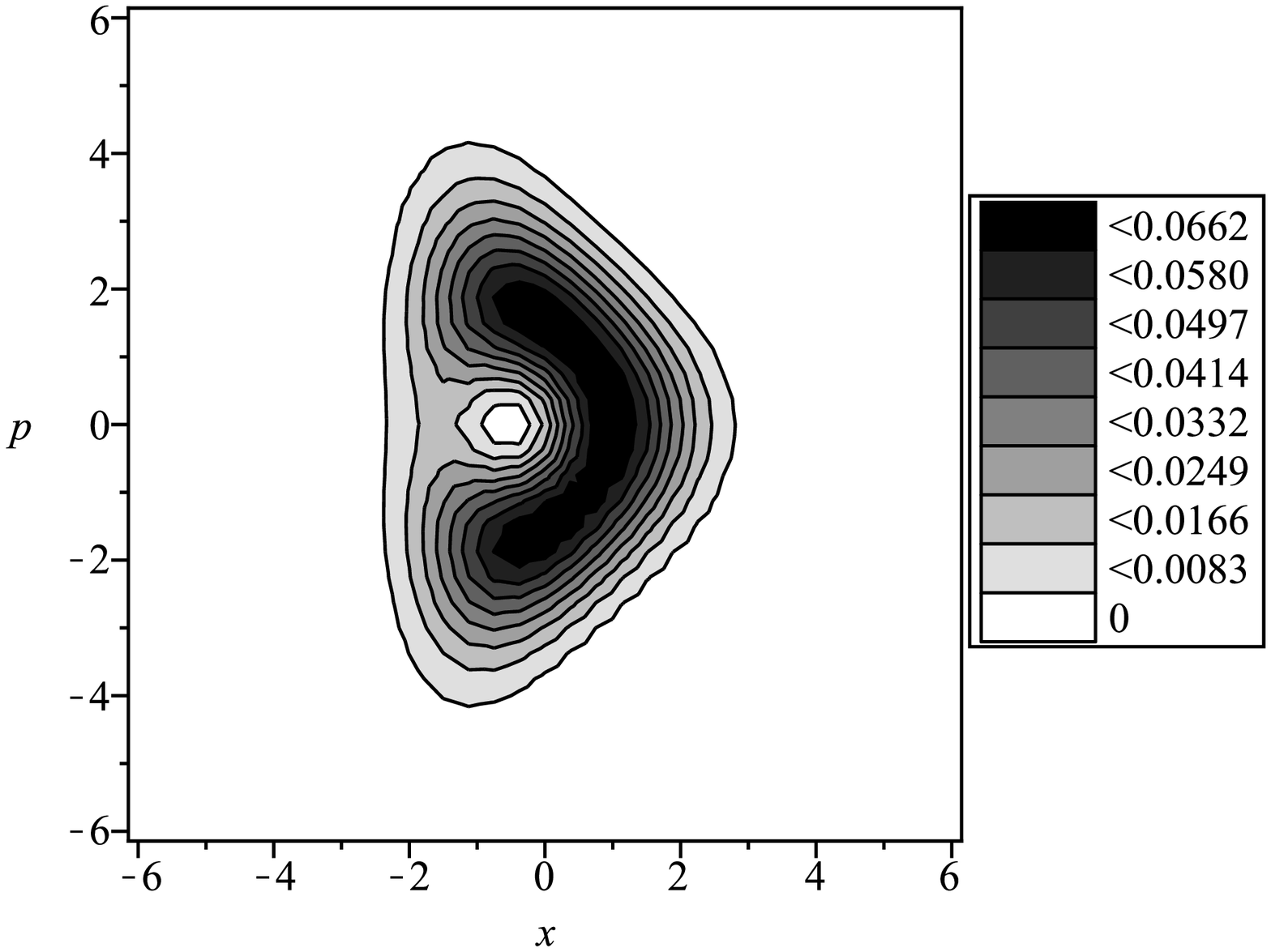}
}\\
\resizebox{0.40\textwidth}{!}{%
  \includegraphics{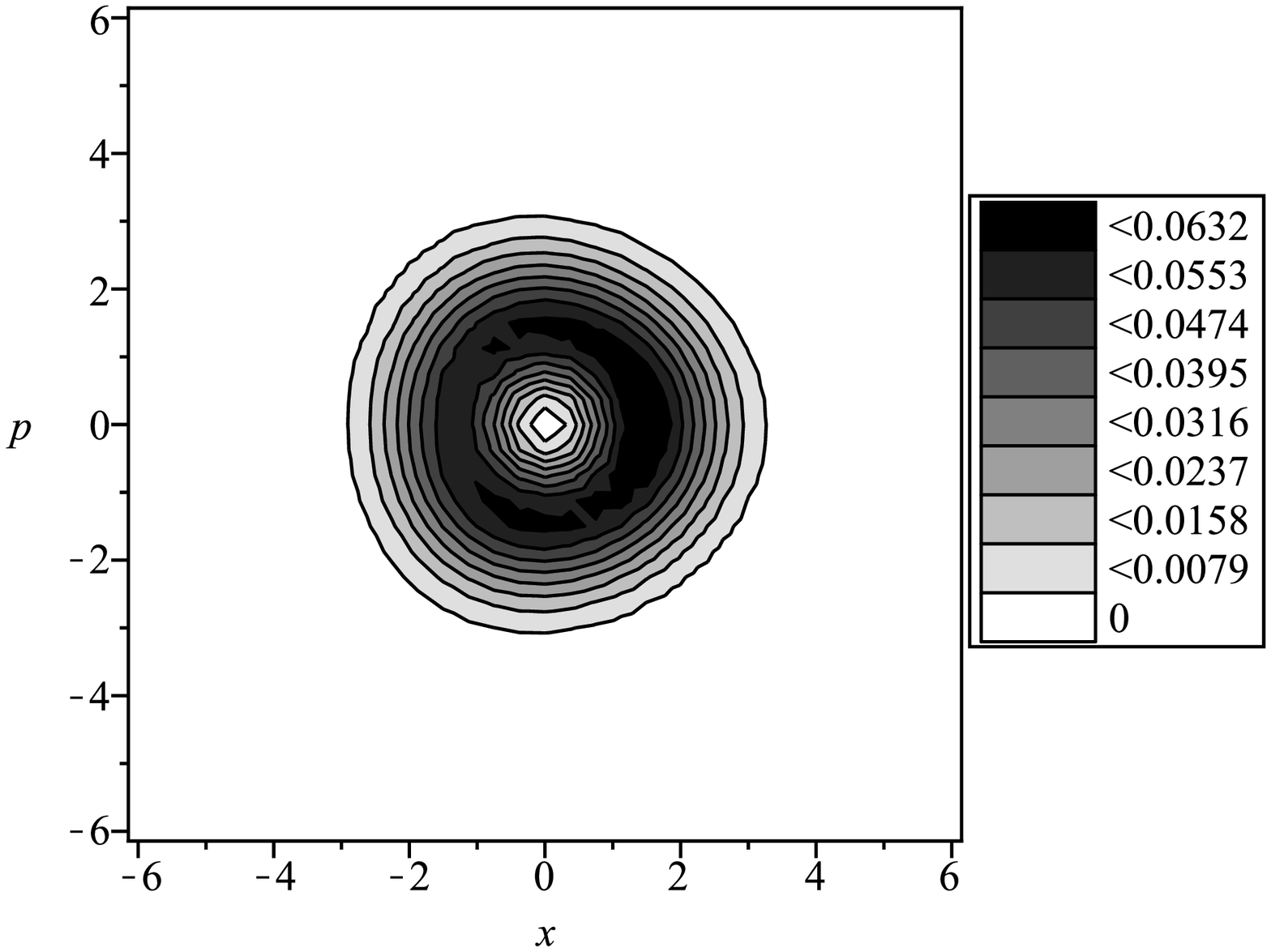}
}
\resizebox{0.40\textwidth}{!}{%
  \includegraphics{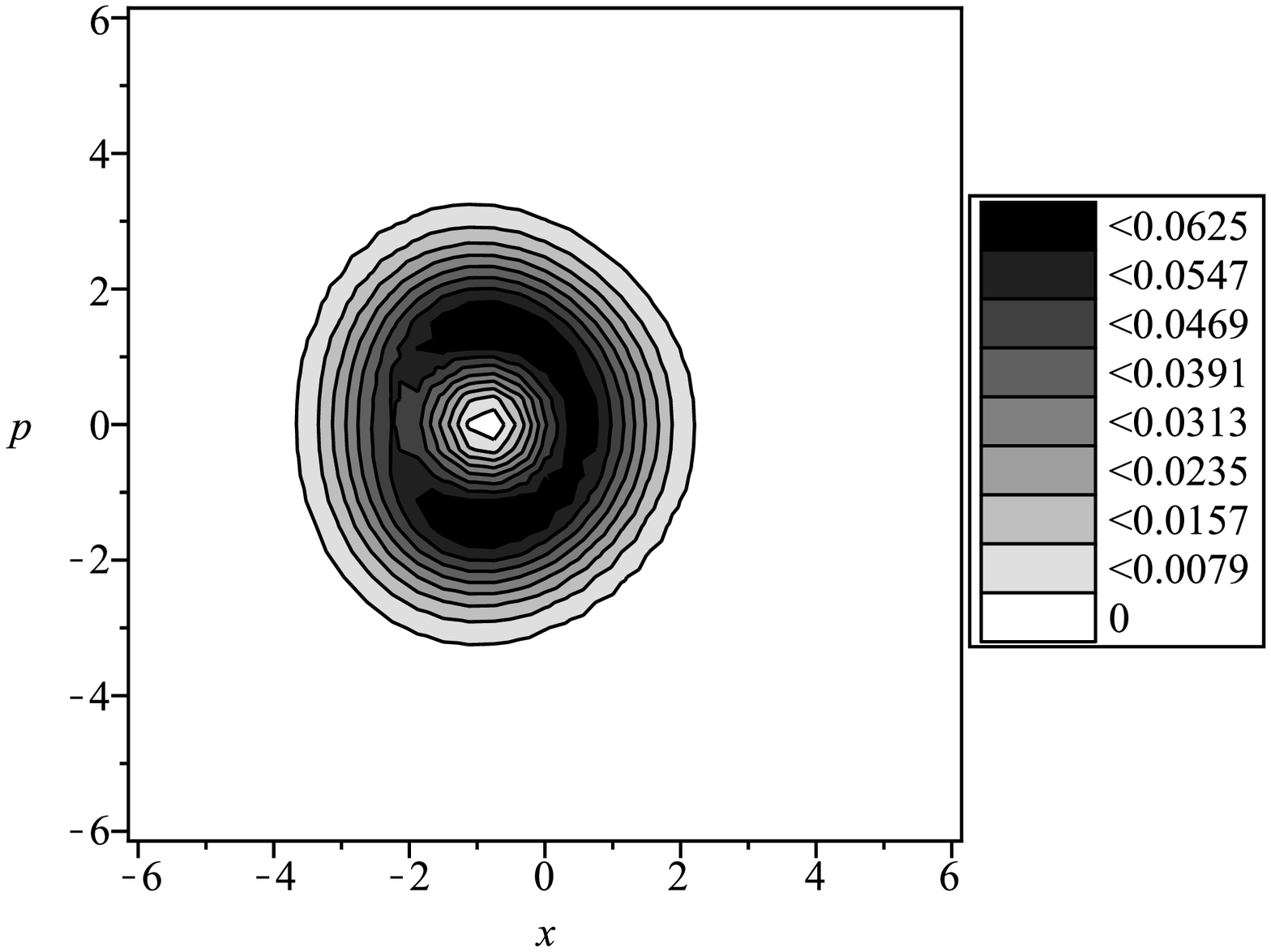}
}
\end{center}
\caption{Comparative plot of the semiconfined quantum harmonic oscillator Husimi function~(\ref{hf-4}) of the first excited state ($n=1$) without an external field ($g=0$, left plots) and with an external field ($g=1$, right plots). Upper plots correspond to the confinement parameter value $a=0.5$, whereas, middle and lower plots correspond to the confinement parameter values $a=2$ and $a=12$ ($m_0=\omega=\hbar=1$).} 
\label{fig.2}
\end{figure}

The first excited state of the model in terms of the Husimi function (shown in Fig.2) exhibits more intricate behavior than the ground state. Here, one observes the behavior that drastically differs from the behavior of the distribution of the probability densities in the $x$-configuration space. The maximum of the joint distribution is in the region corresponding to the value of momentum $p=0$ and position $x>0$. Applied external field $g \neq 0$ slightly smooths this maximum to the positive and negative values of the momentum. 

\cite{zhang2003} highlights that an external field applied to the asymmetrical quantum system adjusts its asymmetry. Such an adjustment under the influence of the stronger external field leads to larger enhancements of the second-order nonlinear optical effects in the asymmetrically fabricated solid-state structures. Mathematically it happens in such a manner due to the following analytical expression of the discrete energy spectrum of the non-relativistic quantum harmonic oscillator under the action of the external homogeneous field $V^{ext}\left(x \right)=gx$:

\be
\label{en-g}
E_n ^g  = \hbar \omega \left( {n + \frac{1}{2}} \right) - \frac{{g^2 }}{{2m_0 \omega ^2 }},\quad n = 0,1,2, \ldots .
\ee

From this expression, it is clear that the applied external field contributes to the second term becomes a reason for the appearance of the second-order nonlinear optical effects in the non-relativistic quantum harmonic oscillator. The case of the asymmetrical potential exhibiting semiconfinement effect only makes the second-order nonlinear optical effects much stronger. Numerical computations show that the intervals between the energy levels decrease and the second-harmonic generation susceptibility becomes larger~\cite{zhang2003}. The energy spectrum of the semiconfinement oscillator model under the present study has the following analytical expression:

\be
\label{e-g}
E_n^{gSC}  = \hbar \omega \sqrt {1 + \frac{{2g}}{{m_0 \omega ^2 a}}} \left( {n + \frac{1}{2} + \frac{{m_0 \omega }}{\hbar }a^2 } \right) - m_0 \omega ^2 a^2  - ag,\quad n = 0,1,2, \ldots .
\ee

\cite{jafarov2022} thoroughly discusses features of this analytical expression, which depends both from the semiconfinement parameter $a$ and external field parameter $g$ as well as from their ratio $g/a$ (or, the parameter $g_0$ introduced via definition (\ref{g0})). The role of this ratio can be directly observed from the exact expression of the Husimi function (\ref{hf-4}) as well as it becomes obvious during limit computations, which will be presented below as well as from the plots presented in both figures. The value of parameter $a$ close to zero can make the second-order nonlinear optical effects stronger even in the case of the applied weaker external field. Such a behavior of the model under study can be observed from the comparison of the left and right upper plots of both figs. 1\&2, where the confinement parameter $a$ equals to $0.5$. The impact of the mentioned above nonlinear effects can be observed through the calculation of the linear, second- and third-harmonic generation susceptibilities for the $Al_xGa_{1-x}As/GaAs$ type heterostructures through the employing the wavefunctions of the semiconfinement oscillator model (\ref{wf-gsc}) and its energy spectrum (\ref{e-g}) under the influence of the external homogeneous field. 

Of course, what we discuss here, are only a few general details of the model behavior in the phase space in terms of the Husimi function. More detailed discussions must also take into account the model's behavior in the presence of a strong external field ($g>>1$) and/or semiconfinement effect close to zero ($a<<\frac 12$). Then, one can observe the picture that will differ drastically from the Gaussian-like one.

The influence of semiconfinement rapidly fades as the value of the semiconfinement parameter $a$ increases, and the triangular-like picture of the joint distribution, which is the consequence of the multiplication of two parabolic cylinder functions, is replaced by a circle-like picture. This picture is evidence of the recovery of the Husimi function (\ref{hf-gh}) of the non-relativistic harmonic oscillator. In other words, drawn figures show that there is a correct limit from the Husimi function (\ref{hf-4}) to (\ref{hf-gh}) under the limit $a\to \infty$. Analytically its proof requires the following straightforward approach, some details of which we provide below.

Let's show that the following correct limit from $\bar Q_0^g$ defined via (\ref{q0g-5}) to $Q_{N0}^g$ defined via (\ref{qn0g}) holds:

\be
\label{lim-q0}
\mathop {\lim }\limits_{b \to \infty } \bar Q_0^g  = Q_{N0}^g .
\ee

We are going to use the following series expansions

\[
\ln \left( {1 + x} \right) \cong x - \frac{{x^2 }}{2},\quad \sqrt {1 + x}  \cong x - \frac{{x^2 }}{8},
\]
and Stirling's approximation

\[
\Gamma \left( {z + 1} \right) \cong \sqrt {2\pi z} e^{z\ln z - z} .
\]

Taking them into account, one can write down that

\[
D_\nu  \left( z \right) \cong \frac{1}{{\sqrt 2 }}\exp \left[ {\frac{\nu }{2}\ln \left( { - \nu } \right) - \frac{\nu }{2} - \sqrt { - \nu } z} \right],
\]
and

\[
g_0  = \sqrt {1 + \frac{{2\xi _0 }}{b}}  \cong 1 + \frac{{\xi _0 ^2 }}{b} - \frac{{\xi _0 ^2 }}{{2b^2 }},\quad \ln \left( {1 + \frac{{2\xi _0 }}{b}} \right) \cong \frac{{2\xi _0 }}{b} - \frac{{2\xi _0 ^2 }}{{b^2 }}.
\]

Next, the following approximations also can be easily obtained:

\begin{eqnarray}
 \left( {2g_0 } \right)^{b^2  + \frac{1}{2}}  &\cong& e^{\delta _1 } ,\quad b^{b^2  + 1}  = e^{\delta _2 } , \nonumber \\ 
 \Gamma \left( {b^2  + 1} \right) &\cong& \sqrt {2\pi } be^{\delta _3 } ,\quad \frac{1}{{\sqrt {\Gamma \left( {2b^2  + 1} \right)} }} \cong \frac{1}{{\sqrt {2b\sqrt \pi  } }}e^{\delta _4 } , \nonumber \\ 
 D_{ - b^2  - 1} \left( z \right) &\cong& \frac{1}{{\sqrt 2 }}e^{\delta _5 } , \nonumber 
\end{eqnarray}
where,

\begin{eqnarray}
 \delta _1  = \left( {b^2  + \frac{1}{2}} \right)\ln 2 + b\xi _0  - \xi _0 ^2 ,\quad \delta _2  = \left( {b^2  + 1} \right)\ln b,\quad \delta _3  = b^2 \ln b^2  - b^2 , \nonumber \\ 
 \delta _4  =  - b^2 \ln \left( {2b^2 } \right) + b^2 ,\quad \delta _5  =  - \left( {b^2  + 1} \right)\ln b + \frac{{b^2 }}{2} - b\left( {\delta  + i\eta } \right) + \frac{1}{2}\xi _0 ^2 . \nonumber 
\end{eqnarray}

By performing trivial computations, one observes that

\[
\delta _1  + \delta _2  + \delta _3  + \delta _4  + \delta _5  + \frac{{z^2 }}{4} + \bar \beta _0  \cong \ln \sqrt 2  - \frac{{\Delta ^2  + \eta ^2 }}{4} + \frac{i}{2}\delta \eta .
\]

Its substitution at (\ref{lim-q0}) proves the correction of that limit relation.

This proof can be extended to the case of the arbitrary $n$, too. Then, the most appropriate way is to show that there is a correct limit of $\bar Q_n^g$ defined via (\ref{qng-5}). Application of the following known limit relation between the Laguerre and Hermite polynomials~\cite{koekoek2010}

\[
\mathop {\lim }\limits_{\alpha  \to \infty } \left( {\frac{2}{\alpha }} \right)^{\frac{1}{2}n} L_n^{\left( \alpha  \right)} \left( {\left( {2\alpha } \right)^{\frac{1}{2}} x + \alpha } \right) = \frac{{\left( { - 1} \right)^n }}{{n!}}H_n \left( x \right)
\]
and further trivial calculations easily prove the correct limit from $\bar Q_n^g$ to $Q_{Nn}^g$ defined via (\ref{qnng-4}).

Finally, we conclude that the phase space representation for a semiconfined harmonic oscillator model with a position-dependent effective mass is constructed in terms of the Gaussian smoothed Wigner function of the joint quasiprobability of momentum and position, which is also called as Husimi function. We have found the exact expression of this distribution function for the stationary states of the oscillator model under consideration for both cases without and with the applied external homogeneous field. It is expressed through the double sum of the parabolic cylinder function and has the correct limit to the Husimi function that corresponds to the so-called Hermite oscillator.

\section*{Acknowledgments}

The authors thank J.~Van der Jeugt for his comments and helpful suggestions during the implementation of this work.

\end{document}